\documentclass[12pt,draftcls,onecolumn]{IEEEtran}

\usepackage{times}

\usepackage{amsthm}
\usepackage{amsmath}    
\usepackage{amssymb}    
\usepackage{mathrsfs}   
\usepackage{epsfig}     
\usepackage{graphicx}
\usepackage{colordvi}
\usepackage{xcolor}
\usepackage{epstopdf,caption}

\newcommand{\cN}{{\cal N}}
\newcommand{\cO}{{\cal O}}

\newcommand{\bA}{\mathbf{A}}
\newcommand{\bB}{\mathbf{B}}
\newcommand{\bC}{\mathbf{C}}

\newcommand{\bH}{\mathbf{H}}
\newcommand{\bI}{\mathbf{I}}

\newcommand{\bK}{\mathbf{K}}

\newcommand{\bP}{\mathbf{P}}
\newcommand{\bQ}{\mathbf{Q}}
\newcommand{\bR}{\mathbf{R}}

\newcommand{\bW}{\mathbf{W}}

\newcommand{\bY}{\mathbf{Y}}

\newcommand{\bq}{\mathbf{q}}
\newcommand{\by}{\mathbf{y}}
\newcommand{\bw}{\mathbf{w}}

\newcommand{\be}{\mathbf{e}}

\newcommand{\bh}{\mathbf{h}}
\newcommand{\bu}{\mathbf{u}}

\newcommand{\bz}{\mathbf{z}}

\renewcommand{\leq}{\leqslant}

\renewcommand{\geq}{\geqslant}

\allowdisplaybreaks

\newcommand{\Cref}[1]{Co\-ro\-lla\-ry\,\ref{#1}}

\newcommand{\Fref}[1]{Figure~\ref{#1}}

\newcommand{\eq}[1]{(\ref{#1})}

\begin{document}


\title{Price of Anarchy in Multiuser Massive MIMO: Coordinated versus Uncoordinated Channel Tracking for High-Rate Internet of Things}



\author{Hediyeh Soltanizadeh, Shahrokh Farahmand, and S. Mohammad Razavizadeh
\thanks{H. Soltanizadeh, S. Farahmand, and S. M. Razavizadeh are with the School
of Electrical Engineering, Iran University of Science and Technology (IUST), Tehran, Iran, e-mail: (he.soltanizadeh@elec.iust.ac.ir,\{shahrokhf,smrazavi\}@iust.ac.ir).}
\thanks{Manuscript last edited \today}}

\markboth{IEEE Transactions on Wireless Communications (submitted)}%
{IEEE Transactions on Wireless Communications (submitted)}

\maketitle



\begin{abstract}
Incorporation of high-rate internet of things (IoT) service into a massive MIMO framework is investigated. It is revealed that massive MIMO possess the inherent potential to offer such service provided it knows the channels for all devices. Our proposed method is to jointly estimate and track the channels of all devices irrespective of their current activity. Using the dynamical model for devices' channels evolution over time, optimal and sub-optimal trackers are developed for coordinated scenario. Furthermore, we introduce a new paradigm where the BS need not know the pilot access patterns of devices in advance which we refer to as uncoordinated setup. After motivating this scenario, we derive the optimal tracker which is intractable. Then, target tracking approaches are applied to address uncertainties in the measurements and derive sub-optimal trackers. Our proposed approaches explicitly address the channel aging problem and will not require downlink paging and uplink access request control channels which can become bottlenecks in crowded scenarios. The fundamental minimum mean square error (MMSE) gap between optimal coordinated and uncoordinated trackers which is defined as price of anarchy is evaluated and upper-bounded. Stability of optimal trackers is also investigated. Finally, performance of various proposed trackers are numerically compared.
\end{abstract}
\IEEEpeerreviewmaketitle

\section{Introduction}
Mobile broadband (MBB) and massive machine-type communication (mMTC) constitute two main application areas in future generations of cellular networks. mMTC is also referred to as the crowded or overloaded scenario and is mainly driven by internet of things (IoT) applications. To accommodate these two type of services in future networks, promising cellular concepts such as massive MIMO should be adapted to their particular characteristics. Recent efforts on fusion of these concepts are divided into three categories: i) MBB, ii) low-rate IoT, iii) high-rate IoT. 

Benefiting from channel hardening property, massive MIMO decouples individual users channels into deterministic ones free from interference, small-scale fading, and noise \cite{M10}, while applying low-complexity linear beamforming/combining techniques \cite{LETM14,spmagazine}. As its major challenge, all users channels should be known at the BS. Orthogonal pilots have been utilized to enable channel estimation \cite{M10}. They were proven to be optimal for peer-to-peer ordinary MIMO as well \cite{gersh}. Each orthogonal pilot should be assigned to one user only to avoid pilot contamination. This is possible for MBB as there are only a few active users. A well-known initial random access procedure for dynamic pilot assignment in MBB is random access to pilots (RAP) which is reminiscent of slotted ALOHA with the difference that collision domain is no longer time but pilot sequences \cite{SCP14}. RAP has been further modified to permit collision resolution \cite{BCSLP17,SAMD16}. 

The proposed grant-based approaches are not suitable for low-rate IoT. Firstly, crowded scenarios lead to excessive collisions in a slotted ALOHA system greatly reducing its success rate. Secondly, every device has only a few bits and the overall access request/grant procedure incurs too much overhead. Thirdly, there might not be even enough orthogonal pilots for all simultaneously active devices. The proposed remedy is to transmit a pilot pattern followed immediately by the few data bits without any previous handshake with the BS. These schemes are referred to as grant-free and are divided into two broad categories depending on the pilot structure. First category advocates non-orthogonal pilot sequences that are assigned uniquely and permanently to every device. The corresponding decoders mostly rely on compressed sensing \cite{beyene17,du17,du18,yu17,mywork}. Second category recommends sharing orthogonal pilots among several users. Ergodic random access to pilots including data (E-RAPiD) and coded RAPiD (C-RAPiD) are two examples \cite{RAsurvey}. In addition to the original C-RAPiD \cite{SCSP16} which borrowed ideas from collision resolution in slotted ALOHA \cite{italians07}, MBB methods such as \cite{BCSLP17} has also been modified to incorporate similar collision resolution procedures \cite{HLG17}.

Proposed approaches for low-rate IoT pose certain limitations in high-rate IoT. Firstly, neither of the proposed methods can handle large data rates. When using non-orthogonal random access, either payload length following a pilot transmission should be greatly enlarged or many successive access slots should be used. Payload enlargement is limited by the channel coherence time, while many successive access attempts will destroy the sparseness of activity patterns, severely limiting the decoder performance. E-RAPiD will incur excessive large decoding delays as long data packets should be communicated over many coherence times. Thus, real-time applications such as voice, video, robotics surgery, and interactive IoT can not be supported. C-RAPiD will suffer from a lower delay, given that it uses one coherence time, but will have to deal with many active users in every coherence time. This translates to decoding coded messages with many loops and active nodes which will adversely affect message-passing techniques. Two other issues were not addressed by any of the above methods. In crowded scenarios, too many devices need to be paged at every coherence time which jams the downlink control channel. It would be desirable to remove this bottleneck. Secondly, all aforementioned methods assume a block fading model which overlooks channel aging. 

To the best of our knowledge, the only two available channel tracking methods for massive MIMO are presented in \cite{DK17} and \cite{KFtrack}. To be specific, \cite{DK17} does not transmit pilots and only data is transmitted. In fact, \cite{DK17} iterates between turbo decoding and channel estimation using previously decoded data as pilots. The proposed approach is limited by the fact
that it does not use dynamical models on the channel and does not benefit
from optimal sequential estimation procedures such as Kalman filter (KF). Furthermore, error propagation occurs if turbo codes fail to correct all errors. On the other hand, \cite{KFtrack} uses pilot-hopping in combination with a KF to track the channel parameters variations. It treats interfering devices as noise and can not benefit from the extra information that their channel estimates will provide. Furthermore, it will not perform satisfactorily if PH pattern and/or pilot transmission times are unknown. 
 
Using a resolvable multipath channel model in the angular domain, an alternative approach to enable pilot sharing was offered in \cite{ZXGJJL18}. The angular spread of each device can be further tracked in time \cite{GDZXDW17}. Two major limitations of this approach are: i) must have no local scatterers near BS, ii) does not exploit the resolution capability of small-scale fading, also known as fast fading, which is a consequence of lack of enough scatterers.

\subsection{Main Contributions}
Massive MIMO offers an inherent potential to address all the aforementioned concerns for high-rate IoT. The challenge is to learn all the devices channels. By explicitly considering a dynamic model for channel evolution, we propose joint trackers that simultaneously follow all devices irrespective of their current activity status. To enable high-quality tracking, we decouple pilot transmission from data communication. Thus, in each access slot, one device may transmit pilot only to help improve its available channel estimate, may transmit data only when BS already maintains a good channel estimate for that user, may transmit both pilot and data, or remain completely silent. We also advocate orthogonal pilot sharing, similar to low-rate IoT, but use the different spatial features of various devices to resolve collisions. Our proposed spatial feature for every user is its best available channel estimate from previous access slots. Unlike large-scale path loss \cite{BCSLP17} or distance to BS \cite{SAMD16} which can be the same for many devices in crowded scenarios, the whole channel vector will be different for various users with very high probability. Subsequently, we will propose several low-complexity sub-optimal trackers as well. Our proposed approaches will i) resolve pilot collisions and exploit their information ii) omit the need for separate downlink paging and uplink control channels, iii) explicitly account for channel aging, iv) offer minimum decoding delays compared to E-RAPiD/C-RAPiD, v) decouple pilot transmission and data communication patterns allowing for more flexibility. Our next major contribution, which has not been considered in prior art, is to address the problem of unknown pilot transmission patterns by various devices at the BS. In Section IV we motivate this scenario and offer the corresponding optimal and sub-optimal trackers. Our cheif contributions are:
\begin{enumerate}
\item[1.] When BS knows pilot transmission patterns, which we refer to as coordinated scenario, optimum joint tracker and two low-complexity sub-optimal trackers are introduced.
\item[2.] When BS does not know pilot transmission patterns of devices, which we refer to as uncoordinated scenario, optimal but intractable tracker is derived. Borrowing ideas from target tracking community, low-complexity sub-optimal trackers are also introduced. To further reduce complexity, three novel heuristic trackers are also derived.
\item[3.] The performance gap between the two optimal trackers, which offers a fundamental performance limit and is referred to as price of anarchy, is evaluated and an upper bound for it is derived. Stability of optimal trackers is also investigated.
\item[4.] Performance of the proposed trackers are investigated through extensive simulations.   
\end{enumerate}

\subsection{Organization}
Section II formulates the problem. Section III presents coordinated channel trackers. Section IV offers the optimal and three sub-optimal uncoordinated trackers. Section V presents the three heuristic low-complexity trackers for the uncoordinated setup. Section VI provides performance analysis. Section VII provides numerical results and Section VIII concludes the paper.


\section{System model and problem formulation}
\label{sec:def}

Let us consider a single cell where a MIMO base station (BS) equipped with $M$ antennas is serving a total of $N$ single-antenna IoT devices. The system operates in a time-division duplex (TDD) mode, where each access slot is divided into 3 segments. In the first segment, a pilot of length $\tau$ is transmitted by some of the IoT devices and BS exploits the received signal to track devices channel gains. In the second segment, uplink data are transmitted followed by downlink data in the third segment. As $N \gg \tau$, there exist more devices than orthogonal pilots. Thus, BS associates the same pilot to a group of $K$ devices where $K:=N/\tau$ is an integer. For Section V.A, we have $K:=N/(\tau-1)$ as \eqref{na} asks for one pilot sequence to be unused. To enable channel tracking, each device transmits the pilot irregularly once in a while according to a random pattern. Different devices may transmit pilots at different rates. In a single random access slot, some devices may transmit data only, some may transmit pilot only, and some transmit both pilot and data while the rest remain totally silent. The objective is to track all IoT devices channels at successive random access slots. Once this challenge is addressed, BS can utilize the obtained channel estimates to perform various beamforming methods at uplink/downlink. It should be noted that BS tracks all devices channels irrespective of whether they have data to transmit or are silent for now. Subsequently, downlink beamforming will be applied whenever downlink data for a particular device is available. On the other hand, uplink beamforming is exploited at every access slot to find those users which have data to transmit to the BS. As the channel tracking process for devices belonging to distinct groups can be decoupled by the orthogonality of their pilots, we focus on channel tracking for users within group one corresponding to pilot sequence one or $\boldsymbol{\phi}_1\in {\bf R}^{\tau\times 1}$ without loss of generality.

We aim to track the flat fading channel gains between the $k$’th device and the BS in the $t$’th time slot, which is represented by ${\bf{h}}_{t}^{(k)} ={[h_{t}^{(k)}(1), h_{t}^{(k)}(2)  \cdots  h_{t}^{(k)}(M)]}^{T}$ for $k=1 \dots K$. Note that $\bh_t^{(k)}$ includes large-scale path-loss, shadowing and small-scale fading. Furthermore, we assume power allocation is also absorbed into $\bh_t^{(k)}$. Note that we implicitly assume a MIMO-OFDM system but focus on one coherence bandwidth. Equivalently, we can assume a narrowband single-carrier system. Instead of adopting a typical block-fading model, we deal with channel aging directly and treat the channels as time-varying dynamical systems. Hence, user $k$ channel gains change from access slot $t-1$ to access slot $t$ according to the given state model: 
\begin{equation}
\label{eq:1}
{\bf{h}}_{t}^{(k)}={\bf{A}}_t^{(k)} {\bf{h}}_{t-1}^{(k)} + {\bu}_{t}^{(k)}
\end{equation}
where $\bA_t^{(k)}$s represent known model matrices that determines how fast or  slow the channel changes over time slots and $\bu_t^{(k)}$ denotes Gaussian process noise with zero mean and covariance matrix $\bQ_t^{(k)}$. Process noise is independent across $t,k$. To simplify matters and without loss of generality, we consider a real system with real channel gains. Extension to the complex case will be straightforward. As for initialization, we assume $\bh_t^{(k)}\sim \cN ({\bf 0},{\bf I}_{M})$. The model in \eqref{eq:1} is fairly general and very flexible. One can check \cite{KFtrack} for certain structures that fit into this model and those that do not. Different users channels evolve independent of one another. This is ensured by the independence of initial channels assignment and independence of process noise for various devices.
 
Each device at a particular access slot decides whether or not to send the pilot with the probability $\lambda_k$ independently from other devices. Thus, only a subset of devices are active and send pilots within each access slot. We define a binary variable $q_t^{(k)}$ that takes one if the $k$'th device transmits its pilot at time slot $t$. Let $D_{t}$ denote the subset of all active devices in time slot $t$ who transmit $\boldsymbol{\phi}_1$ with cardinality $|D_t|=N_a$. Then, the received signal at the BS at time slot $t$ is given by $\bY_t\in{\bf R}^{M\times\tau}$:
\begin{equation}
\label{eq:2}
{\bY_t}=\sum_{k=1}^{K}q_t^{(k)}{\bh_t^{(k)}}\boldsymbol{\phi}_1^T+{\bW_t}
\end{equation}
where $\bW_t \in \mathbb{R}^{M \times \tau}$ indicates the additive white Gaussian noise (AWGN) matrix whose entries are independent with zero mean and variance equal $\sigma_w^2$. Given $\bY_{1:t}$, our objective is to estimate $\bh_t^{(1:K)}$ which amounts to the channels of all users who are allocated to group one. If $q_t^{(k)}$'s are known to the BS, the tracking scheme is coordinated. However, when $q_t^{(k)}$'s are not known at BS, we refer to the scheme as uncoordinated. Our performance criterion is minimum mean square error (MMSE). All of our proposed trackers assume an accurate initial acquisition of all devices channels at access slot one. This can be achieved by separating devices in different groups via orthogonal pilots and devices within a group by TDMA to perform an initial training. While extensive, acquisition is performed only once, hence it does not have a significant effect on the overall performance of the trackers.  Trackers update all users channel estimates at the pilot transmission phase of each access slot. At the uplink/downlink stages, BS uses the latest channel estimates to perform beamforming towards various devices. Next section deals with coordinated channel tracking where three algorithms with varying degrees of complexity-performance trade-off are presented.

\section{Tracking with Coordinated Pilot Access}
First, we provide the optimum tracker which is a Kalman filter that tracks an aggregated state containing the joint state of all the $K$ devices. Then, two sub-optimal alternatives with lower complexity are offered. We refer to all three as coordinated because BS knows the pilot transmission pattern $q_t^{(k)}$s.

\subsection{Joint Coordinated Kalman Filter (JC-KF)}
While \eqref{eq:1} suggests that channels for different devices evolve independently and thus independent individual trackers might be optimal, the optimum tracker can not be decoupled across devices. This comes naturally as measurements in \eqref{eq:2} do include collisions which introduce coupling between various devices. First, we remove the effect of the devices in other groups by multiplying measurements with the orthonormal pilot of group 1:
\begin{equation} \label{meas_model}
\by_t:=\bY_t\boldsymbol{\phi}_1=\sum_{k=1}^{K}q_t^{(k)}{\bh_t^{(k)}}+{\bw_t}.
\end{equation} 
Here, $\by_t\in {\mathbb R}^{M\times 1}$ represents the measurement vector corresponding to group one received at the $M$ antennas at time $t$ and $\bw_t:=\bW_t\boldsymbol{\phi}_1$ denotes the corresponding noise vector which is still Gaussian and independent across entries with variance $\sigma_w^2$. We use the general covariance $\bR_t$ in place of $\sigma_w^2\bI$ to allow for correlated measurement noise also. Note that measurement noises should be independent over time. Upon introducing the aggregate state vector $\bh_t:=[\bh_t^{(1)^T},\bh_t^{(2)^T},\ldots,\bh_t^{(K)^T}]^T$ of size $MK$, $\by_t$ can be written as
\begin{equation} \label{meas_model2}
\by_t=\left[\begin{array}{cccc}q_t^{(1)}\bI_M~\vline & q_t^{(2)}\bI_M~\vline &\cdots \hspace{0.2cm}\vline &q_t^{(K)}\bI_M\end{array}\right]\bh_t+\bw_t:=\bB_t\bh_t+\bw_t.
\end{equation}
Furthermore, \eqref{eq:1} can be written in a joint state format as
\begin{equation}\label{state_model}
\bh_t=\bA_t\bh_{t-1}+\bu_t,
\end{equation}
where $\bA_t$ of size $KM\times KM$ is block diagonal with $\bA_t^{(k)}$s on the diagonals and $\bu_t:=[\bu_t^{(1)^T},\bu_t^{(2)^T},\ldots,\bu_t^{(K)^T}]^T$ is zero-mean with a block diagonal covariance matrix $\bQ_t$ with $\bQ_t^{(k)}$s on the diagonal. Note that state equation in \eqref{state_model} is linear with all variables being jointly Gaussian. When $\bB_t$s are known, measurement model is also linear with all variables being jointly Gaussian. Therefore, the optimum MMSE tracker is a Kalman filter run on the aggregate state. As in a typical KF, JC-KF begins with an initial estimate obtained from acquisition stage and then iteratively performs prediction and correction steps \cite{K93}.\\
{\bf{Prediction:}}
\begin{equation}
\label{eq:5}
{\bf{\hat{h}}}_{t|t-1}={\bf{A}}_t {\bf{\hat{h}}}_{t-1|t-1},\quad  {\bf{P}}_{t|t-1}={\bf{A}}_t {\bf{P}}_{t-1|t-1} {\bf{A}}_t^{T}+\bQ_t.
\end{equation}
{\bf Correction:} First define ${\bf{K}}_{t}={\bf{P}}_{t|t-1}\bB_t^T\left(\bB_t{\bf{P}}_{t|t-1}\bB_t^T+\bR_t\right)^{-1}$. Then,
\begin{equation}
{\bf{\hat{h}}}_{t|t}={\bf{\hat{h}}}_{t|t-1}+{\bf{K}}_{t}({\by_t}-\bB_t{\bf{\hat{h}}}_{t|t-1}),\qquad  \bP_{t|t}=\left(\bI-{\bf{K}}_{t}\bB_t\right){\bf{P}}_{t|t-1}.
\end{equation}
The JC-KF serves as a benchmark to compare all other tracking algorithms that will be introduced later on, including coordinated and uncoordinated methods, as it offers the smallest possible MMSE. Before proceeding two remarks are in order.\\
{\bf Remark 1.} The covariance matrix $\bP_{t|t}$ is of size $KM \times KM$ and contains the covariance matrix for channel estimate of device $k$ on its $k$'th diagonal block. Furthermore, it contains the cross-covariance between users $k$ and $j$ channel estimates on the $(k,j)$'th block.\\
{\bf Remark 2.} Depending on the various devices activity patterns that forms $\bB_t$ in \eqref{meas_model2}, the MMSE for user $k$ can be different from user $j$. Note that MMSE for user $k$ is given by the trace of $\bP_{k,k}$ which is the $k$'th $M\times M$ block on the diagonal of $\bP_{t|t}$. This MMSE is independent of the measurements, and only a function of $\bB_t$. It provides the fundamental MMSE limit that is achievable by the joint activity pattern encoded in $\bB_t$.  

\subsection{Coordinated Kalman Filter with Collisions Discarded}
The chief limitation of JC-KF lies in its high complexity as its aggregated state is of size $MK$ and the corresponding covariance matrix is if size $(MK)^2$. For large $K$, JC-KF can become prohibitively complex. To address this challenge we introduce two sub-optimal alternatives which utilize independent Kalman filters for various devices. Note that if we discard the $\by_t$ whenever it suffers a collision, coupling is removed. Therefore, we introduce coordinated independent KF (CI-KF) as follows. For each device $k$, run the following iterations independently.\\
{\bf{Prediction:}}
\begin{equation}
\label{eq:5}
{\bf{\hat{h}}}_{t|t-1}^{(k)}={\bf{A}}_t^{(k)} {\bf{\hat{h}}}_{t-1|t-1}^{(k)},\quad  {\bf{P}}_{t|t-1}^{(k)}={\bf{A}}_t^{(k)} {\bf{P}}_{t-1|t-1}^{(k)} {\bf{A}}_t^{{(k)}^T}+\bQ_t^{(k)}.
\end{equation}
{\bf Correction:} If only user $k$ transmitted $\boldsymbol{\phi}_1$ perform correction as below. Otherwise, skip to the prediction for next access slot. For correction, define ${\bf{K}}_{t}={\bf{P}}_{t|t-1}^{(k)}\left({\bf{P}}_{t|t-1}^{(k)}+\bR_t\right)^{-1}$. Then,
\begin{equation}
{\bf{\hat{h}}}_{t|t}^{(k)}={\bf{\hat{h}}}_{t|t-1}^{(k)}+{\bf{K}}_{t}({\by_t}-{\bf{\hat{h}}}_{t|t-1}^{(k)}),\qquad  \bP_{t|t}^{(k)}=\left(\bI-{\bf{K}}_{t}\right){\bf{P}}_{t|t-1}^{(k)}.
\end{equation}
If there is a collision, the corresponding $\by_t$ is discarded and only prediction is performed for all devices. If only a single device accessed the pilot, the channel estimate corresponding to that user is both predicted and corrected, while other devices channel estimates are only predicted. CI-KF operates satisfactorily particularly when number of collisions are a few. This occurs either if $K$ is small or if $K$ is large but BS schedules users so that collisions are minimized. However, CI-KF MMSE is lower bounded by that of JC-KF since CI-KF can not exploit the information contained in the collisions. CI-KF needs $\cO(KM^3)$ arithmetic operations per access slot while JC-KF needs $\cO(K^3M^3)$. Complexity gap is considerable when $K$ is large. 

\subsection{Coordinated Kalman Filter via Belief Propagation}
CI-KF will skip many measurements when collisions are abound and will perform poorly. To overcome this limitation, we introduce a second sub-optimal KF that is derived using belief propagation (BP) on a factor graph. This BP-based KF (BP-KF), exploits collisions while its complexity is of the same order as CI-KF. BP-KF algorithm is concisely presented here, however its derivation is relegated to the Appendix A. 
\begin{enumerate}
	\item 
	Fix $\bh_0^{(k)}~\sim \cN(\hat{\bh}_0^{(k)},\bP_{0|0}^{(k)})$ for all $k=1,\ldots,K$ which are obtained from the initial acquisition stage. Set $t=0$.
	\item At time instant $t\geq 1$ perform the following steps:
	\item 
	Perform prediction step independently for all devices. This amounts to calculating   ${\bf{\hat{h}}}_{t|t-1}^{(k)}=\bA_t^{(k)}\hat{\bh}_{t-1|t-1}^{(k)}$ and $\bP_{t|t-1}^{(k)}=\bA_t^{(k)}\bP_{t|t-1}^{(k)}\bA_t^{(k)^T}+\bQ_t^{(k)}$.
	\item 
	For those users $k$ who are present in the current collision form the fictitious measurement $\hat{\by}_t^{(k)}$ and its covariance matrix $\hat{\bR}_t^{(k)}$ given as
	\begin{equation}\label{fictitious}
	\hat{\by}_t^{(k)}:=\by_t-\sum_{j\neq k}q_t^{(j)}\hat{\bh}_{t|t-1}^{(j)}, \qquad \hat{\bR}_t^{(k)}:=\bR_t+\sum_{j\neq k}q_t^{(j)}\hat{\bP}_{t|t-1}^{(j)}
	\end{equation}
	\item
	Correct the channel estimate for the users that participate in the collision as follows. For user $k$ assume the measurement model $\hat{\by}_t^{(k)}:=\bh_t^{(k)}+\hat{\bw}_t^{(k)}$ where $\hat{\bw}_t^{(k)}$ is the fictitious noise with zero-mean and covariance $\hat{\bR}_t^{(k)}$, then run the correction step. Define ${\bf{K}}_{t}^{(k)}={\bf{P}}_{t|t-1}^{(k)}\left({\bf{P}}_{t|t-1}^{(k)}+\hat{\bR}_t^{(k)}\right)^{-1}$. Then, perform
	\begin{equation} \label{mybp}
	{\bf{\hat{h}}}_{t|t}^{(k)}={\bf{\hat{h}}}_{t|t-1}^{(k)}+{\bf{K}}_{t}^{(k)}({\hat{\by}_t}-{\bf{\hat{h}}}_{t|t-1}^{(k)}),\qquad  \bP_{t|t}=\left(\bI-{\bf{K}}_{t}^{(k)}\right){\bf{P}}_{t|t-1}^{(k)}.
	\end{equation}
	\item 
	Set $t\leftarrow t+1$ and go to step 2.
\end{enumerate}
It is interesting to observe how BP-KF deals with collisions. Firstly, BP-KF amounts to applying JC-KF but after each iteration it discards the cross-correlation terms which are the off diagonal blocks in the joint covariance matrix. This renders the covariance matrix block diagonal leading to independent Kalman filters across users. Second observation is that the aforementioned fictitious measurement compensates for other users presence in the collision by subtracting the best available estimate of their channels but enlarges the covariance of the fictitious measurement noise to account for errors in other users channel estimates which results in \eqref{fictitious}. BP-KF complexity per access slot is $\cO(KM^3)$.

\subsection{Comparison with Prior Art}
Comparing \cite{KFtrack} against our methods, several conclusions should be drawn. First, \cite{KFtrack} treats the desired user channel as an FIR filter in time and tracks all the channel coefficients allowing for tracking the complete channel response in the frequency domain, while we track one coherence bandwidth only. Secondly, \cite{KFtrack} allows for $\bA_t^{(k)}$s to be unknown, while we treat them as known. On the other hand, \cite{KFtrack} fails to utilize the information obtained from tracking other devices as JC-KF does. Indeed, if there are no collisions at all, our CI-KF and tracker in \cite{KFtrack} will be the same. It should be noted that \cite{KFtrack} knows the pilot access pattern of the desired device, hence it amounts to a coordinated approach according to our notation in spite of the term uncoordinated in their title. Comparisons will be performed in the simulations.

\section{Tracking with Uncoordinated Pilot Access}
As its main advantage, coordination allows for a simple tracker with polynomial complexity such as JC-KF to become the MMSE estimate. On the negative side, it diminishes system flexibility. To enable coordination, devices should either use a fixed random pilot access pattern, or periodically notify the BS of changes in access pattern. Changing, or adaptive, access pattern can be necessary from several perspectives. For devices that run on a battery or harvest energy, a fixed pilot access pattern is not justified as they should adapt pilot access rate to their available energy levels. Furthermore, a single device might transmit data with various reliability requirements over time. To increase reliability, the device may decide to increase its pilot access rate, while decreasing it when data is tolerant to errors. However, periodic notifications on new access policies from all IoT devices can jam the control channel and lead to a control bottleneck. To ensure flexibility in pilot access while avoiding excessive control signaling, we offer a novel viewpoint not considered before. The main idea is to allow all devices to individually choose their pilot access slots at will without any coordination with the BS. However, BS will be tasked with the additional burden of detecting pilot access patterns. Therefore, we assume $q_t^{(k)}$s are not known to the BS.
   
As BS does not know in advance which devices are present in a collision, the tracking task becomes significantly more complicated. This challenge has existed for a long time in target tracking community and is referred to as the measurement origin uncertainty problem, the data association problem, or data assignment problem. While the MMSE optimal tracker is prohibitively complex, many sub-optimal  trackers have been developed. Unfortunately, target tracking solutions can not be readily applied to our framework as tracking community always dealt with measurement origin uncertainty, but they always assumed that every single measurement is at most generated by one target. This assumption is no longer valid in our scenario as a combination of devices are present in a collision. We correspondingly modify the available target tracking algorithms to accommodate this new assumption. In this Section, we first present the optimal uncoordinated tracker. Then, we present three sub-optimal trackers which are global nearest neighbor (GNN) \cite{pdaf}, probabilistic data association filter (PDAF) \cite{pdaf}, and multiple hypotheses tracker (MHT) \cite{mht}. 

\subsection{Optimal Tracker}
The optimal, in the MMSE sense, joint tracker is given by
\begin{equation} \label{optMMSE}
\hat{\bh}_t:={\rm E}\left[\bh_t|\by_{1:t}\right]={\rm E}\left[{\rm E}\left[\bh_t|\by_{1:t},\bq_{1:t}\right]|\by_{1:t}\right]=\sum_{\bq_{1:t}} p(\bq_{1:t}|\by_{1:t})~{\rm E}\left[\bh_t|\by_{1:t},\bq_{1:t}\right]
\end{equation}
Conditioned on $\bq_{1:t}$, the inner expected value is simply evaluated by a joint KF that assumes $\bB_{1:t}$s are given according to that specific $\bq_{1:t}$ pattern. To compute the outer expectation, we should enumerate over all possible instances of $\bq_{1:t}$. Note that each $\bq_t$ can assume $2^K$ values and $\bq_{1:t}$ assumes $2^{tK}$ different values. Thus, to evaluate optimal MMSE tracker, $2^{Kt}$ joint KFs should be run in parallel and then combined by the weights $p(\bq_{1:t}|\by_{1:t})$. Indeed, the optimal MMSE tracker is distributed as a Gaussian mixture with exponentially increasing number of mixtures over time. The weights $p(\bq_{1:t}|\by_{1:t})$ are evaluated as follows:
\begin{eqnarray}
p(\bq_{1:t}|\by_{1:t})&=&\frac{p(\by_{1:t}|\bq_{1:t})p(\bq_{1:t})}{p(\by_{1:t})},\qquad p(\bq_{1:t})=\prod_{i=1}^t p(\bq_i)=\prod_{i=1}^t \prod_{k=1}^K \lambda_k^{q_i^{(k)}} (1-\lambda_k)^{1-q_i^{(k)}}\label{opt_w1}\nonumber \\ 
p(\by_{1:t}|\bq_{1:t})&=&p(\by_t|\by_{1:t-1},\bq_{1:t})p(\by_{1:t-1}|\bq_{1:t})=p(\by_t|\by_{1:t-1},\bq_{1:t})p(\by_{1:t-1}|\bq_{1:t-1})\nonumber\\ &&\hspace{-2cm}=\prod_{i=1}^t~p(\by_i|\by_{1:i-1},\bq_{1:i}) =\prod_{i=1}^t~\cN\left(\by_i;\bB_i(\bq_i)\hat{\bh}_{i|i-1},\bB_i(\bq_i)\bP_{i|i-1}\bB_i^T(\bq_i)+\bR_i\right).\label{ekf}
\end{eqnarray}
Note that in \eqref{ekf}, we run a joint KF assuming a given $\bq_{1:t}$ and evaluate how good the given measurements are predicted by this choice of $\bq_{1:t}$. This process is repeated for all possible choices of $\bq_{1:t}$. Once weights are obtained from \eqref{opt_w1}, they are normalized to one to account for the unknown $p(\by_{1:t})$ in the denominator in \eqref{opt_w1} and then plugged into \eqref{optMMSE} to compute the mean and covariance matrix for $\hat{\bh}_t$. The overall complexity of optimal tracker is $\cO(K^3M^32^{Kt})$. 

\subsection{Global Nearest Neighbor}
Optimal tracker enumerates all possible hypotheses on the origin of measurements and then assigns a corresponding weight to each hypothesis. The weight is assigned according to how well that particular hypothesis is predicted by the measurements. Unfortunately, number of hypotheses grow exponentially in $t$ preventing the applicability of the optimal tracker. One simple remedy is to greedily pick the hypothesis that predicts the data best at the current time and fix it for future access slots. This algorithm is referred to as GNN. 

GNN operates as follows. At time $t$, it is assumed that $\bq_{1:t-1}$ are selected correctly. Thus, one only needs to select the best possible $\bq_t$ which assumes $2^K$ values. Unlike the optimal tracker, GNN performs hard assignment meaning that it discards all hypotheses except the best greedy one. GNN algorithm is briefly described below. 
\begin{enumerate}
	\item 
	Fix $\bh_0~\sim \cN(\hat{\bh}_0,\bP_{0|0})$ which are obtained from the initial acquisition stage. Set $t=0$.
	\item At time instant $t\geq 1$ perform the following steps:
	\item 
	Perform prediction step on the joint state. This amounts to calculating   ${\bf{\hat{h}}}_{t|t-1}=\bA_t\hat{\bh}_{t-1|t-1}$ and $\bP_{t|t-1}=\bA_t\bP_{t|t-1}\bA_t^T+\bQ_t$.
	\item 
	Consider all possible hypotheses on $\bq_t$. Then find the one that maximizes $p(\bq_t|\bq_{1:t-1}^\ast,\by_{1:t})$ and denote it by $\bq_t^\ast$. This is done as follows:
	\begin{eqnarray*}
	p(\bq_t|\bq_{1:t-1}^\ast,\by_{1:t})&=&\frac{p(\by_t|\by_{1:t-1},\bq_{1:t-1}^\ast,\bq_t)p(\bq_t|\by_{1:t-1},\bq_{1:t-1}^\ast)}{p(\by_t|\by_{1:t-1},\bq_{1:t-1}^\ast)}\nonumber \\ & &\hspace{-2cm}\propto p(\by_t|\by_{1:t-1},\bq_{1:t-1}^\ast,\bq_t) p(\bq_t)\nonumber \\ & &\hspace{-2cm} \propto  \cN\left(\by_t;\bB_t(\bq_t)\hat{\bh}_{t|t-1},\bB_t(\bq_t)\bP_{t|t-1}\bB_t^T(\bq_t)+\bR_t\right) \prod_{k=1}^K \lambda_k^{q_t^{(k)}} (1-\lambda_k)^{q_t^{(k)}}.
 	\end{eqnarray*}
 	Note that $\hat{\bh}_{t|t-1},\bP_{t|t-1}$ are computed from the previous iterations of the single KF that assumes $\bq_{1:t-1}^\ast$. 
	\item
	Set $\bB_t^\ast:=\bB_t(\bq_t^\ast)$, then run the correction step assuming $\bB_t^\ast$ was the true model. First, define ${\bf{K}}_{t}={\bf{P}}_{t|t-1}\bB_t^{\ast^T}\left(\bB_t^\ast{\bf{P}}_{t|t-1}\bB_t^{\ast^T}+\bR_t\right)^{-1}$. Then, perform
	\begin{equation}
	{\bf{\hat{h}}}_{t|t}={\bf{\hat{h}}}_{t|t-1}+{\bf{K}}_{t}({\by_t}-\bB_t^\ast{\bf{\hat{h}}}_{t|t-1}),\qquad  \bP_{t|t}=\left(\bI-{\bf{K}}_{t}\bB_t^\ast\right){\bf{P}}_{t|t-1}.
	\end{equation}
	\item 
	Set $t\leftarrow t+1$ and go to step 2.
\end{enumerate}
GNN tracker assumes a Gaussian distribution for $\hat{\bh}_{t|t}$ to simplify calculations in contrast to the optimal tracker which is distributed as GMM. Furthermore, it uses a single-step greedy approach to select the best possible $\bq_t$. Once the greedy optimum is chosen it becomes fixed for future random access slots and its optimality is never re-evaluated again.

\subsection{Multiple Hypothesis Tracker}
GNN keeps track of only one hypothesis over time, while the optimal tracker keeps account of all possible hypotheses. To fill the gap between these two, one might suggest to keep track of a fixed $N_h$ hypotheses instead of one. This idea will give rise to MHT. To elaborate, assume $\Omega_{t-1}^{i}:=\left\{\bq_1(i),\bq_2(i),\ldots,\bq_{t-1}(i)\right\}$ denotes the $i$'th hypothesis that is kept by MHT at time $t-1$. The MHT is briefly described below.
\begin{enumerate}
	\item 
	Fix $\bh_0~\sim \cN(\hat{\bh}_0,\bP_{0|0})$ which are obtained from the initial acquisition stage. Set $t=0$. Set $\Omega_0^{i}=\emptyset$ for all $i=1,2,\ldots,N_h$. Set $p(\Omega_0^{i})=1/N_h$ for all $i$.
	\item At time instant $t\geq 1$ perform the following steps:
	\item 
	Perform prediction step on the joint state for all hypotheses. This amounts to calculating   ${\bf{\hat{h}}}_{t|t-1}^{(\Omega_{t-1}^i)}=\bA_t\hat{\bh}_{t-1|t-1}^{(\Omega_{t-1}^i)}$ and $\bP_{t|t-1}^{(\Omega_{t-1}^i)}=\bA_t\bP_{t|t-1}^{(\Omega_{t-1}^i)}\bA_t^T+\bQ_t$ for the $i$'th hypothesis $\Omega_{t-1}^i$.
	\item 
	For every $i$, augment $\Omega_{t-1}^i$ with all possible choices on $\bq_t$. Since $\bq_t$ can take $2^K$ different values, each $\Omega_{t-1}^i$ is expanded into $2^K$ hypotheses. Mathematically, $\tilde{\Omega}_t^{(i,j)}:=\left\{\Omega_{t-1}^i,\bq_t(j)\right\}$ where $j=1,2,\ldots,2^K$ and $i=1,2,\ldots,N_h$. 
	\item 
	Next, evaluate the probability of each hypotheses as follows. 
\begin{eqnarray} \label{hyp}
	\hspace{-1cm}p(\tilde{\Omega}_t^{(i,j)}|\by_{1:t})&=&\frac{p(\by_t|\tilde{\Omega}_t^{(i,j)},\by_{1:t-1})~p(\tilde{\Omega}_t^{(i,j)}|\by_{1:t-1})}{p(\by_t|\by_{1:t-1})}\\ & &\hspace{2.4cm}\hspace{-3cm} \propto \cN\left(\by_t;\bA_t\hat{\bh}_{t|t-1}^{(\Omega_{t-1}^{i})},\bB_t(\bq_t(j))\bP_{t|t-1}^{(\Omega_{t-1}^{i})}\bB_t^T(\bq_t(j))+\bR_t\right)~p(\Omega_{t-1}^i|\by_{1:t-1})\nonumber
	\end{eqnarray}
	\item 
	Select the $N_h$ largest values in \eqref{hyp} among $N_h2^K$ hypotheses and discard the rest. Then, assign these $N_h$ hypotheses to $\Omega_t^1,\Omega_t^2,\ldots,\Omega_t^{N_h}$. 
	\item 
	Normalize the weights as
	\[
	 p(\Omega_{t}^{\hat{i}}|\by_{1:t})=\frac{p(\tilde{\Omega}_t^{(i,j)}|\by_{1:t})}{\sum_{i=1}^{N_h}~p(\tilde{\Omega}_t^{(i,j)}|\by_{1:t})}
	\] 
	where right hand side probabilities are given by \eqref{hyp} when $\Omega_t^{\hat{i}}=\tilde{\Omega}_t^{(i,j)}$ is the $\hat{i}$'th largest probable hypothesis.
	\item 
	For $\hat{i}=1,2,\ldots,N_h$ run the correction step as follows
	$$
	\hspace{-4.1cm}	\bK_t^{(\Omega_t^{\hat{i}})}:=\bP_{t|t-1}^{(\Omega_{t-1}^i)}\bB_t^T(\bq_t(j))\left(\bB_t(\bq_t(j))\bP_{t|t-1}^{(\Omega_{t-1}^i)}\bB_t^T(\bq_t(j))+\bR_t\right)^{-1}
		$$
		$$
	{\bf{\hat{h}}}_{t|t}^{(\Omega_t^{\hat{i}})}={\bf{\hat{h}}}_{t|t-1}^{(\Omega_{t-1}^i)}+{\bf{K}}_{t}^{(\Omega_t^{\hat{i}})}\left({\by_t}-\bB_t(\bq_t(j)){\bf{\hat{h}}}_{t|t-1}^{(\Omega_{t-1}^i)}\right),\quad  \bP_{t|t}^{(\Omega_t^{\hat{i}})}=\left(\bI-{\bf{K}}_{t}^{(\Omega_t^{\hat{i}})}\bB_t(\bq_t(j))\right){\bf{P}}_{t|t-1}^{(\Omega_{t-1}^i)}.
	$$
	\item
	As the estimate at time $t$, select the hypothesis with the largest weight and assign the corresponding ${\bf{\hat{h}}}_{t|t}^{(\Omega_t^{\hat{i}})}$ as MHT tracker output.
	\item
	Set $t\leftarrow t+1$ and go to step 2.
\end{enumerate}
When $N_h=1$, MHT reduces to GNN. Choice of $N_h$ provides a trade-off between complexity and performance. Like GNN, MHT relies on hard assignments as it discards all the unfavorable hypotheses.

\subsection{Probabilistic Data Association Filter}
A characteristic of optimal tracker which is missing in GNN and MHT is its soft (probabilistic) assignment of hypotheses. Therefore, all hypotheses do have an impact on the final estimate but unfavorable ones have a smaller effect than favorable ones. The impact of each hypotheses is determined by its corresponding weight. PDAF utilizes the same idea of soft assignment but does so only for the current measurement. To elaborate further, remember that MMSE is given by $\hat{\bh}_{t|t}={\rm E}[\bh_t|\by_{1:t}]$ and the corresponding probability density function is evaluated as follow:
\begin{eqnarray}\label{pdaf1}
p(\bh_t|\by_{1:t})=\sum_{\bq_t}~p(\bh_t,\bq_t|\by_{1:t})=\sum_{\bq_t}~p(\bq_t|\by_{1:t})~p(\bh_t|\bq_t,\by_{1:t})
\end{eqnarray}
Note that we have conditioned on $\bq_t$ only as opposed to the optimal tracker which conditioned on $\bq_{1:t}$. The summation is over all $2^K$ possible hypotheses on $\bq_t$. The weights on the right hand side of \eqref{pdaf1} can be written as 
\begin{eqnarray} \label{pdaf2}
p(\bq_t|\by_{1:t})=\frac{p(\by_t|\bq_t,\by_{1:t-1})~p(\bq_t|\by_{1:t-1})}{p(\by_t|\by_{1:t-1})} \propto p(\by_t|\bq_t,\by_{1:t-1})~\lambda_k^{q_t^{(k)}}
(1-\lambda_k)^{(1-q_t^{(k)})}
\end{eqnarray}
The first term on the right hand side of \eqref{pdaf2} is further expanded as
\begin{eqnarray}\label{pdaf4}
p(\by_t|\bq_t,\by_{1:t-1})&=&\int p(\by_t|\bh_t,\bq_t,\by_{1:t-1}) ~p(\bh_t|\bq_t,\by_{1:t-1}) ~d\bh_t \label{pdaf3} \\&&\hspace{-1.8cm}=\int p(\by_t|\bh_t,\bq_t) ~p(\bh_t|\by_{1:t-1}) ~d\bh_t\approx \cN\left(\by_t;\bB_t(\bq_t)\hat{\bh}_{t|t-1},\bB_t(\bq_t)\bP_{t|t-1}\bB_t^T(\bq_t)+\bR_t\right) \nonumber 
\end{eqnarray}
The first equality in \eqref{pdaf4} is valid because given $\bh_t,\bq_t$, measurement $\by_t$ is independent of past measurements $\by_{1:t-1}$. In addition, given only past measurements $\by_{1:t-1}$, $\bq_t$ is independent of $\bh_t$. The approximation occurs because $p(\bh_t|\by_{1:t-1})$ is given by the optimal tracker which is too complex and thus we approximate it using the predictions on the previous step of PDAF. Upon setting $p(\bh_t|\by_{1:t-1})\approx\cN(\bh_t;\hat{\bh}_{t|t-1},\bP_{t|t-1})$ and $p(\by_t|\bh_t,\bq_t)=\cN(\by_t;\bB_t(\bq_t)\bh_t,\bR_t)$ and integrating over $\bh_t$ the approximation ensues. Same approximation is utilized in evaluating the second term in \eqref{pdaf1}. Specifically,
\begin{eqnarray}\label{pdaf5}
&&\hspace{-2.3cm}p(\bh_t|\bq_t,\by_{1:t})=\frac{p(\by_t|\bh_t,\bq_t,\by_{1:t-1})~p(\bh_t|\bq_t,\by_{1:t-1})}{p(\by_t|\bq_t,\by_{1:t-1})} \propto p(\by_t|\bh_t,\bq_t)~p(\bh_t|\by_{1:t-1}) \\ &&\approx \cN(\by_t;\bB_t(\bq_t)\bh_t,\bR_t)~ \cN(\bh_t;\hat{\bh}_{t|t-1},\bP_{t|t-1})\propto \cN\left(\bh_t;\hat{\bh}_{t|t}(\bq_t),\bP_{t|t}(\bq_t))\right)\nonumber 
\end{eqnarray}
where 
\begin{eqnarray*}
\bK_t(\bq_t)&=&\bP_{t|t-1}\bB_t^T(\bq_t)\left(\bB_t(\bq_t)\bP_{t|t-1}\bB_t^T(\bq_t)+\bR_t\right)^{-1}\nonumber\\
\hat{\bh}_{t|t}(\bq_t)&=&\hat{\bh}_{t|t-1}+\bK_t(\bq_t)(\by_t-\bB_t(\bq_t)\hat{\bh}_{t|t-1}),\quad \bP_{t|t}(\bq_t)=\left(\bI-\bK_t(\bq_t)\bB_t(\bq_t)\right)\bP_{t|t-1}
\end{eqnarray*}
$\bK_t(\bq_t)$ denotes the Kalman gain assuming the hypothesis $\bq_t$ is true. Similarly, $\hat{\bh}_{t|t}(\bq_t),\bP_{t|t}(\bq_t)$ are the corresponding mean and covariance after correction step conditioned on $\bq_t$. Note that \eqref{pdaf5} illustrates an alternative derivation of the correction step for the Kalman filter. To summarize, the weights in \eqref{pdaf1} are evaluated via \eqref{pdaf2} and \eqref{pdaf4} for all possible $2^K$ hypotheses over $\bq_t$ and normalized to one. We also plug the approximation in \eqref{pdaf5} into \eqref{pdaf1} and will arrive at a Gaussian mixture. PDAF estimate of mean and covariance is given by the Gaussian mixture in \eqref{pdaf1}. It is straightforward to show that this mean and covariance are given by
\begin{eqnarray}\label{pdaf6}
\hat{\bh}_{t|t}=\sum_{\bq_t}p(\bq_t|\by_{1:t})\hat{\bh}_{t|t}(\bq_t),\quad \bP_{t|t}=\sum_{\bq_t}p(\bq_t|\by_{1:t})\left(\bP_{t|t}(\bq_t)+\hat{\bh}_{t|t}(\bq_t)\hat{\bh}_{t|t}^T(\bq_t)\right)-\hat{\bh}_{t|t}\hat{\bh}_{t|t}^T
\end{eqnarray}
PDAF is briefly described as follows.
\begin{enumerate}
	\item 
	Fix $\bh_0~\sim \cN(\hat{\bh}_0,\bP_{0|0})$ which are obtained from the initial acquisition stage. Set $t=0$.
	\item At time instant $t\geq 1$ perform the following steps:
	\item 
	Perform prediction step on the joint state. This amounts to calculating   ${\bf{\hat{h}}}_{t|t-1}=\bA_t\hat{\bh}_{t-1|t-1}$ and $\bP_{t|t-1}=\bA_t\bP_{t|t-1}\bA_t^T+\bQ_t$.
	\item 
	Consider all possible $2^K$ hypotheses on $\bq_t$. Compute the weights as in \eqref{pdaf1} using \eqref{pdaf2},\eqref{pdaf4}. Then, compute the $2^K$ correction steps for different $\bq_t$ according to \eqref{pdaf5}.
	\item 
	Compute the mean and covariance of PDAF according to \eqref{pdaf6}.
	\item 
	Set $t\leftarrow t+1$ and go to step 2. 
\end{enumerate}

\section{Low-Complexity Channel Tracking with Uncoordinated Pilot Access}
While linear in $t$, GNN, MHT, and PDAF complexities are exponential in $K$ limiting the values of $K$ for which these algorithms are tractable. This section offers three heuristic remedies. The first two maintain polynomial complexity while the third has a random complexity whose worst case can be exponential, however its average performance was similar to the first two methods as verified by numerical comparisons. 

\subsection{Discarding Collisions}
The first solution is to discard collisions as we had done in the coordinated access case. According to \cite{HLG19}, the number of colliding users can be estimated in massive MIMO via
\begin{equation}\label{na}
\hat{N}_a:=\frac{\|\bY_t\boldsymbol{\phi}_1\|_2^2-\|\bY_t\boldsymbol{\phi}_e\|_2^2}{M},
\end{equation}
where $\boldsymbol{\phi}_e$ is an unused orthonormal pilot. For large $M$, law of large numbers ensures convergence of \eqref{na} to the number of colliding users. This is the case only if ${\rm E}[(h_t^{(k)}(m))^2]=1$ for all $t,k,m$ values, that is power control is applied. Upon discarding collisions when $\hat{N}_a \geq 2$, we are left with a possible choice of $K+1$ hypotheses which amounts to only one element of $\bq_t$ being equal to one or all being equal to zero. We can apply any of the three methods of GNN, MHT, and PDAF with $\bq_t$ assuming $K+1$ hypotheses instead of the original $2^K$. This remedy performs satisfactorily only when collisions are a few. Therefore, either $K$ should be small to avoid many collisions or there should be some sort of orthogonality in access patterns which requires extra coordination. Still for small $K=2,3,4$ this scheme can serve four times as many devices as that of dedicated orthogonal pilot assignment. 

\subsection{Soft and Hard Least Squares}
We can rewrite \eqref{meas_model} as follows:
\begin{equation} \label{alt_meas_model}
\by_t=\bH_t\bq_t+\bw_t
\end{equation}
where $\bH_t:=[\bh_t^{(1)}~|\bh_t^{(2)}~|\cdots|~\bh_t^{(K)}]$. If $\bH_t$ was known, we could have estimated $\bq_t$ via an integer constrained least-squares (LS). Subsequently, two difficulties arise. First, integer constraints on $\bq_t$ make the problem significantly harder to solve. Secondly, we do not know $\bH_t$. To address these challenges, we apply two approximation. First, we relax the integer constraint $\bq_t\in\{0,1\}^K$ into a convex hypercube $\bq_t\in[0,1]^K$. Furthermore, we replace the unknown $\bH_t$ with its best available estimate $\hat{\bH}_t:=[\hat{\bh}_{t|t-1}^{(1)}~|\hat{\bh}_{t|t-1}^{(2)}~|\cdots|~\hat{\bh}_{t|t-1}^{(K)}]$. Upon utilizing these approximations, unconstrained LS is immediately obtained as
\begin{equation}
\label{eq:18}
\hat{\bq}_t=\left(\hat{\bH}_t^{T}\hat{\bH}_t\right)^{-1}\hat{\bH}_t^{T}\by_{t}
\end{equation}
For soft LS, we project the obtained $\hat{\bq}_t$ into the hypercube $[0,1]^K$ which simply amounts to rounding values of $\hat{\bq}_t$ out of the $[0,1]$ interval to either $0$ or $1$ whichever is closer and keeping the values inside the interval intact. Given the convex quadratic cost for LS, this optimization then projection approach is equivalent to solving the constrained LS with a hypercube constraint.

For hard LS, we project the obtained $\hat{\bq}_t$ into the set $\{0,1\}^K$ which amounts to rounding each value to either $0$ or $1$. Note that hard LS makes hard decisions on if a particular user is present or absent in a collision while soft LS weighs each user corresponding to its unconstrained LS estimate.  

\subsection{Locally Optimum Maximum Likelihood}
While very simple complexity-wise, the soft/hard LS algorithms in the previous section might perform poorly due to their underlying approximations. One can improve their performance by evaluating the maximum-likelihood estimate (MLE) instead of LS. The joint density of $\by_{1:t}$ parameterized by $\bq_t$ is written as
\begin{equation} \label{bayes}
p\left(\by_{1:t};\bq_t\right)=p\left(\by_{t}|\by_{1:t-1};\bq_t\right)p\left(\by_{1:t-1};\bq_t\right)=p\left(\by_{t}|\by_{1:t-1};\bq_t\right)p\left(\by_{1:t-1}\right)
\end{equation}
The second equality follows because $\by_{1:t-1}$ does not depend on $\bq_t$. To maximize the joint density over $\bq_t$, we should maximize $p\left(\by_{t}|\by_{1:t-1};\bq_t\right)$ which is distributed as GMM (remember the optimum uncoordinated filter). To ensure tractability, we assume that the MLE filter has correctly found $\bq_{1:t-1}$. This assumption has been made by all the other sub-optimal filters of Sections IV and V as well. Then, one can write \eqref{alt_meas_model} as
\begin{equation} \label{alt_meas_model2}
\by_t=\bH_t\bq_t+\bw_t=\hat{\bH}_t\bq_t+\underset{\be_t}{\underbrace{(\bH_t-\hat{\bH}_t)\bq_t+\bw_t}}
\end{equation}
where given $\by_{1:t-1}$, the first term yields the mean and the second term, which is $\be_t$, yields the zero-mean Gaussian noise with its covariance given by
\begin{align}
&\bC_{\be\be}=
\mathbb{E}\left[\be_t\be_t^{T}|\by_{1:t-1}\right]=\mathbb{E}\left\{ \left[(\bH_t-\hat{\bH}_t)\bq_t+\bw_t\right]\left[(\bH_t-\hat{\bH}_t)\bq_t+\bw_t\right]^{T}\right\} \nonumber \\
&\hspace{0.7cm}=\mathbb{E}\left\{ \left[\sum_{k=1}^K q_t^{(k)}(\bh_t^{(k)}-\hat{\bh}_{t|t-1}^{(k)})+\bw_t\right]\left[\sum_{\ell=1}^K q_t^{(\ell)}(\bh_t^{(\ell)}-\hat{\bh}_{t|t-1}^{(\ell)})+\bw_t\right]^{T}\right\}
\nonumber \\
&\hspace{0.7cm}=\sum_{k=1}^K \sum_{\ell=1}^K q_t^{(k)} q_t^{(\ell)} \left(\bP_{t|t-1}\right)_{k,\ell}+\bR_t \nonumber
\end{align}
where $\left(\bP_{t|t-1}\right)_{k,\ell}$ corresponds to the $k,\ell$ block of the joint covariance. MLE maximizes the log-likelihood given by
\begin{equation}
\label{mle}
\hat{\bq}_t=\arg ~ \underset{\bq_t\in\{0,1\}^K}{\max}~-\frac{1}{2}\log |\bC_{\be\be}|-\frac{1}{2}(\by_{t}-\hat{\bH}_t\bq_t)^{T}\bC_{\be\be}^{-1}(\by_{t}-\hat{\bH}_t\bq_t)
\end{equation}
Due to the integer constraints, MLE is difficult to compute. We apply coordinate ascent (CA). Given that each CA step improves ML objective, which is bounded above, its convergence to a local optimum is guaranteed. We begin with $\bq_t={\bf 0}$ and each time vary a coordinate $k$, which is $q_t^{(k)}$, between zero and one with all the other coordinates fixed. Then, select the choice that yields a higher objective. Then, we move on to the next coordinate. In the worst-case the algorithm converges in $2^K$ steps. Finally, we should note that, if we added a prior on $\bq_t$ in \eqref{bayes} we arrived at an sub-optimal maximum a-posteriori (MAP) estimate which yields the same computational complexity as MLE. 

\section{Performance Analysis}
We will examine the MMSE difference between the optimal coordinated versus uncoordinated filters which offers the fundamental performance limit. This gap is referred to as price of anarchy (PoA) because it determines the increased MMSE when users change pilot access patterns at will and do not notify the BS. For an ordinary KF, covariance update is independent of measurements and can be carried offline. Proceeding with the JC-KF, we have
\begin{align} \label{cmmse}
\text{MMSE}_c&=\mathbb{E}\left[\|\bh_{t}-\hat{\bh}^{(c)}_{t|t}\|^{2}\right]=\mathbb{E}\left[\mathbb{E}\left[\mathbb{E}\left[\|\bh_{t}-\hat{\bh}^{(c)}_{t|t}\|^{2}| \bq_{1:t},\by_{1:t}\right]|\bq_{1:t}\right]\right]\nonumber \\ &=\mathbb{E}\left[\mathbb{E}\left[\mbox{trace}(\bP_{t|t}(\bq_{1:t}))|\bq_{1:t}\right]\right]=\sum_{\bq_{1:t}} p(\bq_{1:t}) ~\mbox{trace}(\bP_{t|t}(\bq_{1:t}))
\end{align} 
The inner most expectation, which is conditioned on $\bq_{1:t},\by_{1:t}$ amounts to JC-KF and its MMSE is easily obtained by the trace of the KF covariance matrix tailored to that particular pattern of $\bq_{1:t}$ and is independent of $\by_t$. Each measurement pattern $\bq_{1:t}$ yields a corresponding MMSE given by $\mbox{trace}(\bP_{t|t}(\bq_{1:t}))$. The overall MMSE, averaged over $\bq_{1:t}$ is given by the last equation.

Let us focus on the uncoordinated optimum tracker and its MMSE:
\begin{align} \label{ummse}
\text{MMSE}_u&=\mathbb{E}\left[\|\bh_{t}-\hat{\bh}^{(u)}_{t|t}\|^{2}\right]=\mathbb{E}\left[\mathbb{E}\left[\|\bh_{t}-\hat{\bh}^{(u)}_{t|t}\|^{2}| \by_{1:t}\right]\right]
\end{align} 
Given $\by_{1:t}$, $\hat{\bh}_{t|t}^{(u)}$ is only a function of measurements and hence non-random. Only $\bh_t$ is random in the inner expectation and distributed as the Gaussian mixture in \eqref{optMMSE}. One can write
\begin{align}
\label{eq:42}
&\hspace{-0.5cm}\mathbb{E}\left[\|\bh_{t}-\hat{\bh}^{(u)}_{t|t}\|^{2}|\by_{1:t}\right]=\int\|\bh_{t}-\hat{\bh}^{(u)}_{t|t}\|^{2}\sum_{\bq_{1:t}}~p(\bq_{1:t}|\by_{1:t})\cN(\bh_{t};\hat{\bh}_{t|t}(\bq_{1:t}),\bP_{t|t}(\bq_{1:t}))~d\bh_{t}\nonumber\\ \nonumber
&\hspace{0.7cm}=\sum_{\bq_{1:t}}~p(\bq_{1:t}|\by_{1:t})\int\|\bh_{t}-\hat{\bh}_{t|t}(\bq_{1:t})+\hat{\bh}_{t|t}(\bq_{1:t})-\hat{\bh}^{(u)}_{t|t}\|^{2}\cN(\bh_{t};\hat{\bh}_{t|t}(\bq_{1:t}),\bP_{t|t}(\bq_{1:t}))~d\bh_{t}\\
&\hspace{0.7cm}=\sum_{\bq_{1:t}}~p(\bq_{1:t}|\by_{1:t})\left[\mbox{trace}(\bP_{t|t}(\bq_{1:t}))+\|\hat{\bh}_{t|t}(\bq_{1:t})-\hat{\bh}^{(u)}_{t|t}\|^2\right]
\end{align}
In going from the second to third line, we have expanded the $\|.\|^2$ and used the fact that we are integrating with respect to $\cN(\bh_{t};\hat{\bh}_{t|t}(\bq_{1:t}),\bP_{t|t}(\bq_{1:t}))$ and hence the term  $\|\bh_{t}-\hat{\bh}_{t|t}(\bq_{1:t})\|^2$ will be given by the trace of the corresponding covariance matrix. The term $\|\hat{\bh}_{t|t}(\bq_{1:t})-\hat{\bh}^{(u)}_{t|t}\|^2$ is constant and the cross-term is zero. Next, we plug \eqref{eq:42} into \eqref{ummse} to obtain the final MMSE
\begin{align} \label{finalummse}
\text{MMSE}_u&=\mathbb{E}_{\by_{1:t}}\left[\sum_{\bq_{1:t}}~p(\bq_{1:t}|\by_{1:t})\left[\mbox{trace}(\bP_{t|t}(\bq_{1:t}))+\|\hat{\bh}_{t|t}(\bq_{1:t})-\hat{\bh}^{(u)}_{t|t}\|^2\right]\right]\nonumber\\ &=\sum_{\bq_{1:t}} p(\bq_{1:t}) ~\mbox{trace}(\bP_{t|t}(\bq_{1:t}))+\mathbb{E}_{\by_{1:t}}\left[\sum_{\bq_{1:t}}~p(\bq_{1:t}|\by_{1:t})\|\hat{\bh}_{t|t}(\bq_{1:t})-\hat{\bh}^{(u)}_{t|t}\|^2\right]
\end{align}
The first term in \eqref{finalummse} equals \eqref{cmmse}. The second term yields MMSE difference between the optimal coordinated and uncoordinated trackers. It is the price of anarchy (PoA). It is desirable to characterize PoA analytically such as proving its boundedness and obtaining an upper bound that tells us how much performance we lose due to lack of coordination. These questions are difficult to address generally. However, for some special cases, they can be answered. First, we derive a single step upper bound on PoA. That is we assume both filters used the same initial estimate $\hat{\bh}_0,\hat{\bP}_{0,0}$ and check the negative effect of anarchy at time-step $t=1$. The following theorem ensues whose proof is relegated to the Appendix B.\\
{\bf Theorem 1.}~Assuming the same initial estimate $\hat{\bh}_0,\hat{\bP}_{0,0}$ for both the coordinated and uncoordinated optimum trackers, the single-step PoA is upper bounded by
\begin{align}\label{eq:thm1}
\mbox{PoA} & \leq \sum_{\bq_{1}} p(\bq_1)\Bigg[\hat{\bh}_{1|0}^{T}{\bB}(\bq_1)^T{\bK}(\bq_1)^{T}{\bK}(\bq_1){\bB}(\bq_1)\hat{\bh}_{1|0}+\mbox{trace}\left(\bR+{\bB}(\bq_1)\bP_{1|0}{\bB}(\bq_1)^T\right) \nonumber \\  &\hspace{2.5cm}-\sum_{\tilde{\bq}_1} P(\tilde{\bq}_1)\hat{\bh}_{1|0}^{T}{\bB}(\bq_1)^T{\bK}(\bq_1)^{T}{\bK}(\tilde{\bq}_1){\bB}(\tilde{\bq}_1)\hat{\bh}_{1|0}\Bigg]~.
\end{align}

\subsection{Mean Square (MS) sense Stability of Optimal Trackers}
Unfortunately, Theorem 1 can not determine if PoA is bounded for large $t$ because the presented upperbound grows unbounded as $t$ increases. First, we consider a dynamical system which is stable. In the time-invariant parameters case, this means that all eigenvalues of $\bA_t:=\bA$ have absolute value less than one. The following theorem ensues. The proofs are trivial \cite{AM79}.\\
{\bf Theorem 2.} When the aggregate (or joint) dynamical system in \eqref{state_model} is time-invariant and stable in the MS sense, following conclusions can be drawn:
\begin{enumerate}
	\item[1.] Covariance matrix for the state remains bounded and converges to the unique positive semi-definite solution of Lyapunov equation $\bP=\bA\bP\bA^T+\bQ$.
	\item[2.] JC-KF is MS sense stable meaning that its covariance matrix remain bounded.
	\item[3.] Optimal uncoordinated tracker is MS sense stable meaning that its covariance matrix remain bounded.
	\item[4.] PoA remains bounded.
\end{enumerate}
When the dynamical system of the state is stable, optimal filters are guaranteed to be stable without the need for any observability condition. Stability is guaranteed by the fact that optimal MMSE tracker will have a smaller trace of covariance than the unobserved dynamical system as exploiting measurements optimally can only improve MMSE. As the original covariance for the unobserved system is bounded so does the MMSE for the optimal tracker \cite{AM79}. Subsequently, PoA which is the difference between MSE of optimal trackers will be bounded as well. 

It is well-known that if a time-varying dynamical system is MS sense unstable but uniformly completely observable and controllable, the corresponding Kalman filter is guaranteed to be stable  \cite{AM79, D91, K14}. This is a sufficient condition but not necessary as weaker detectability condition can be used instead. Our final theorem yields sufficient conditions for the stability of JC-KF.\\
{\bf Theorem 3.} If the dynamical system in \eqref{state_model} is MS sense unstable but uniformly completely controllable, then JC-KF is MS sense stable if the prior $p(\bq_{1:t})$ is non-zero only for those $\bq_{1:t}$ combinations which maintain uniform complete observability.

{\it Proof:} Results trivially from \eqref{cmmse}.\\ 
Given the condition in Thm. 3, the optimal coordinated tracker is stable. It is of interest to determine the stability or lack of it for the optimal uncoordinated tracker under the same conditions. This will lead to boundedness / unboundedness conclusion for PoA when observability/controllability conditions hold. However, this is a challenging task and demands further investigation on its own. We leave it as an open problem to be addressed in future works. 

\section{Numerical Results}
We pursue four distinct goals in our simulations. First, we draw a comparison between coordinated and uncoordinated schemes to determine PoA numerically. Secondly, we check the effect of number of antennas on performance to decide how an ordinary MIMO fairs against massive MIMO. Third, we check the data rates that can be achieved via the proposed tracking schemes. Finally, we compare against existing alternatives. Certain parameters are fixed for all simulations. They include $\tau = 16$, $T = 200$, $\rho=0.95$, $\bQ_t^{(k)}=(1-\rho^2){\bf I}_M$, and $\bA_t^{(k)}=\rho {\bf{I}}_{M}$ for all $t,k$. Furthermore, we set $\bR_t={\bf I}_M$, $\lambda_k=(K-1)/K$ for all $k=1,2,\ldots,K$. While we consider setups with $K=2,6$ devices, we only plot the results pertaining to first device as all parameters are selected symmetrically and at random for all devices. We normalize the channel mean-square error (MSE) for all the trackers by the trace of the covariance matrix $(\bP_{t|t})_{1,1}$ which corresponds to the optimal coordinated tracker given by JC-KF. 
\subsection{Performance of Coordinated Methods}
\begin{figure}[t]
	\centering
	
	\begin{minipage}{.5\textwidth}
		\centering
		\captionsetup{font=small, width=.8\linewidth}
		\includegraphics[width=.9\linewidth]{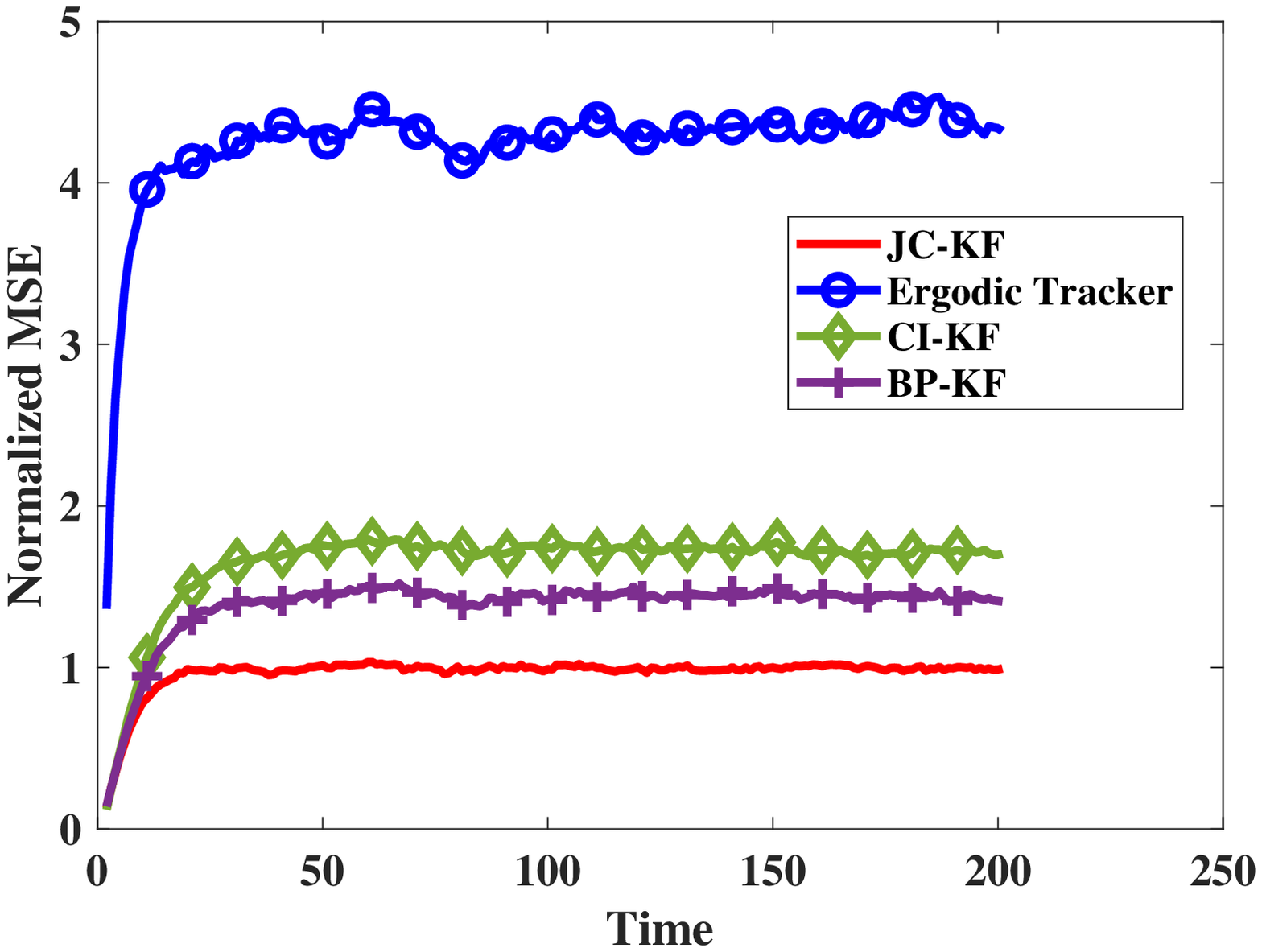}
		\caption{Normalized MSE of coordinated trackers with  $K=6$, $M=16$.}
		\label{fig:coordinateK6M16}
	\end{minipage}%
	\begin{minipage}{.5\textwidth}
		\centering
		\captionsetup{font=small, width=.8\linewidth}
		\includegraphics[width=.9\linewidth]{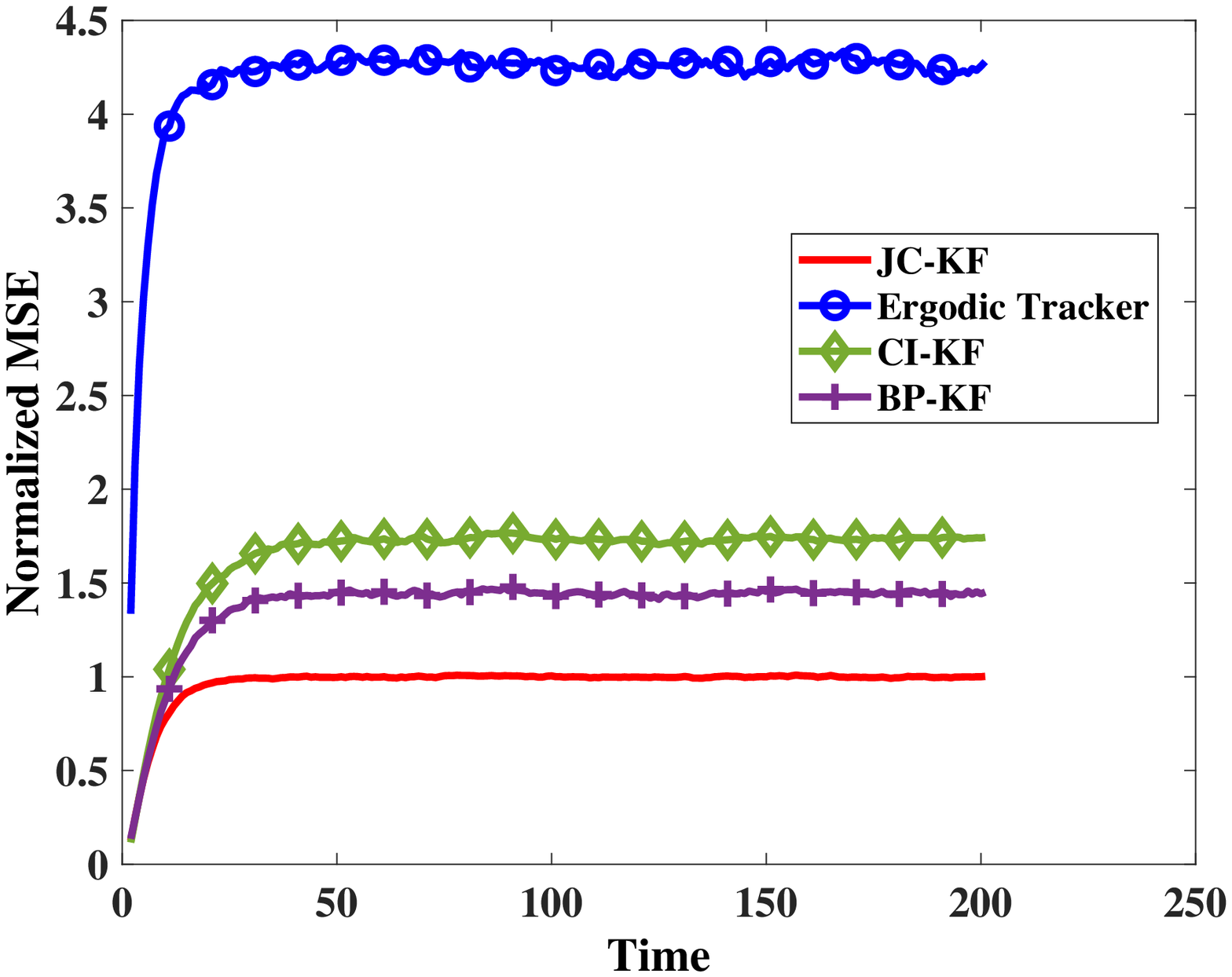}
		\caption{Normalized MSE of coordinated trackers with $K=6$, $M=256$.}
		\label{fig:coordinateK6M256}
	\end{minipage}
\end{figure}

We consider $K=6$ and plot the normalized MSE versus time-slot in  \Fref{fig:coordinateK6M16}, and \Fref{fig:coordinateK6M256} for $M=16$ and $M=256$ antennas respectively. These figures depict a comparison of the JC-KF, CI-KF, and BP-KF. As expected, JC-KF performs best with a normalized MSE of one. For this setup, BP-KF performs slightly better than CI-KF as CI-KF discards too many collisions and thus keeps only predicting all the time. However, the pattern reverses for $K=2$ where collisions are fewer and CI-KF outperforms BP-KF. Due to space limitation, we have omitted the relevant figures for $K=2$. Ergodic tracker which is the algorithm proposed by \cite{KFtrack} performs poorly in both ordinary and massive MIMO scenarios.

\subsection{Performance of Uncoordinated Methods}
\begin{figure}[t]
	\centering
	
	\begin{minipage}{.5\textwidth}
		\centering
		\captionsetup{font=small, width=.8\linewidth}
		\includegraphics[width=.9\linewidth]{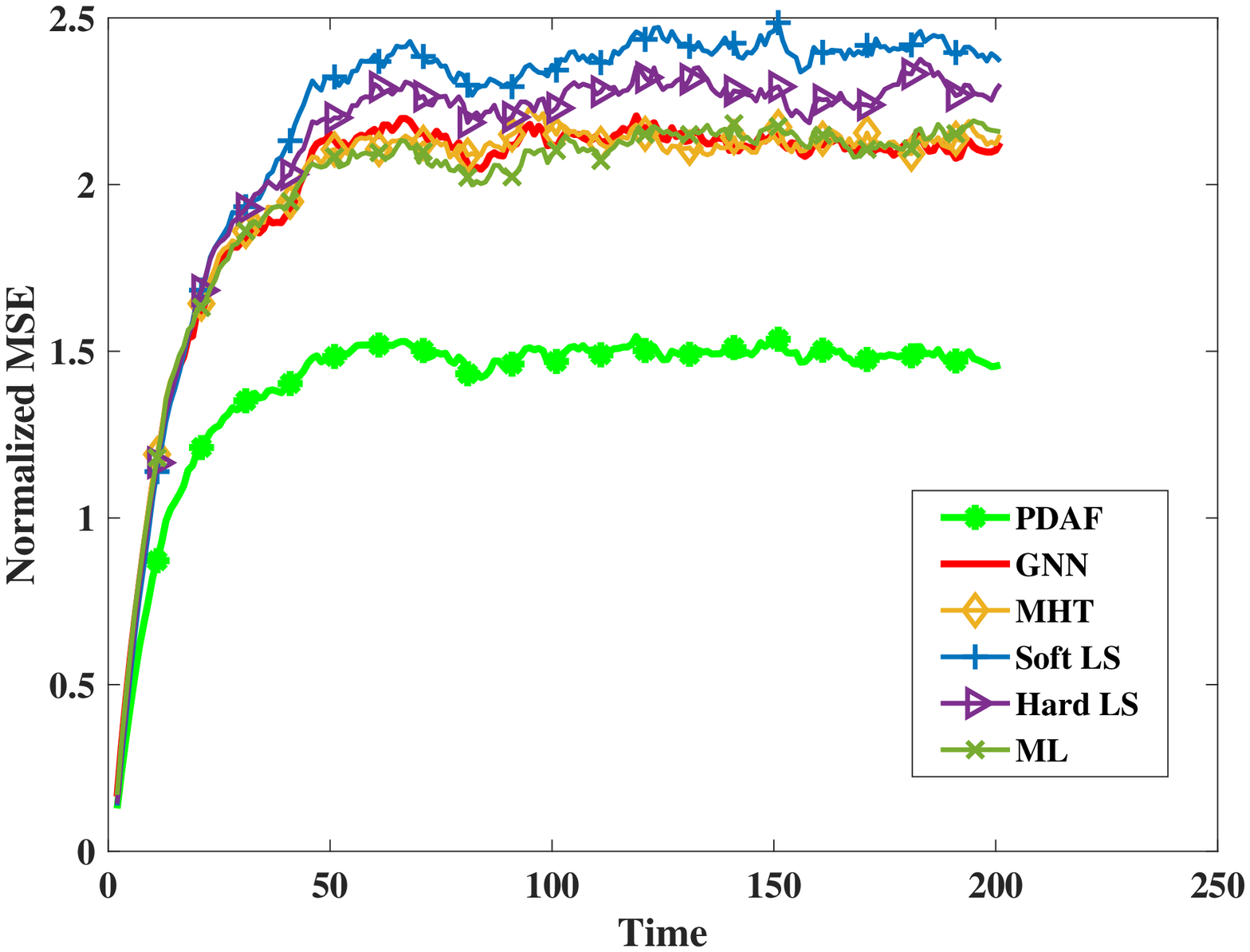}
		\caption{Normalized MSE of uncoordinated trackers with  $K=6$, $M=16$.}
		\label{fig:exploitK6M16}
	\end{minipage}%
	\begin{minipage}{.5\textwidth}
		\centering
		\captionsetup{font=small, width=.8\linewidth}
		\includegraphics[width=.9\linewidth]{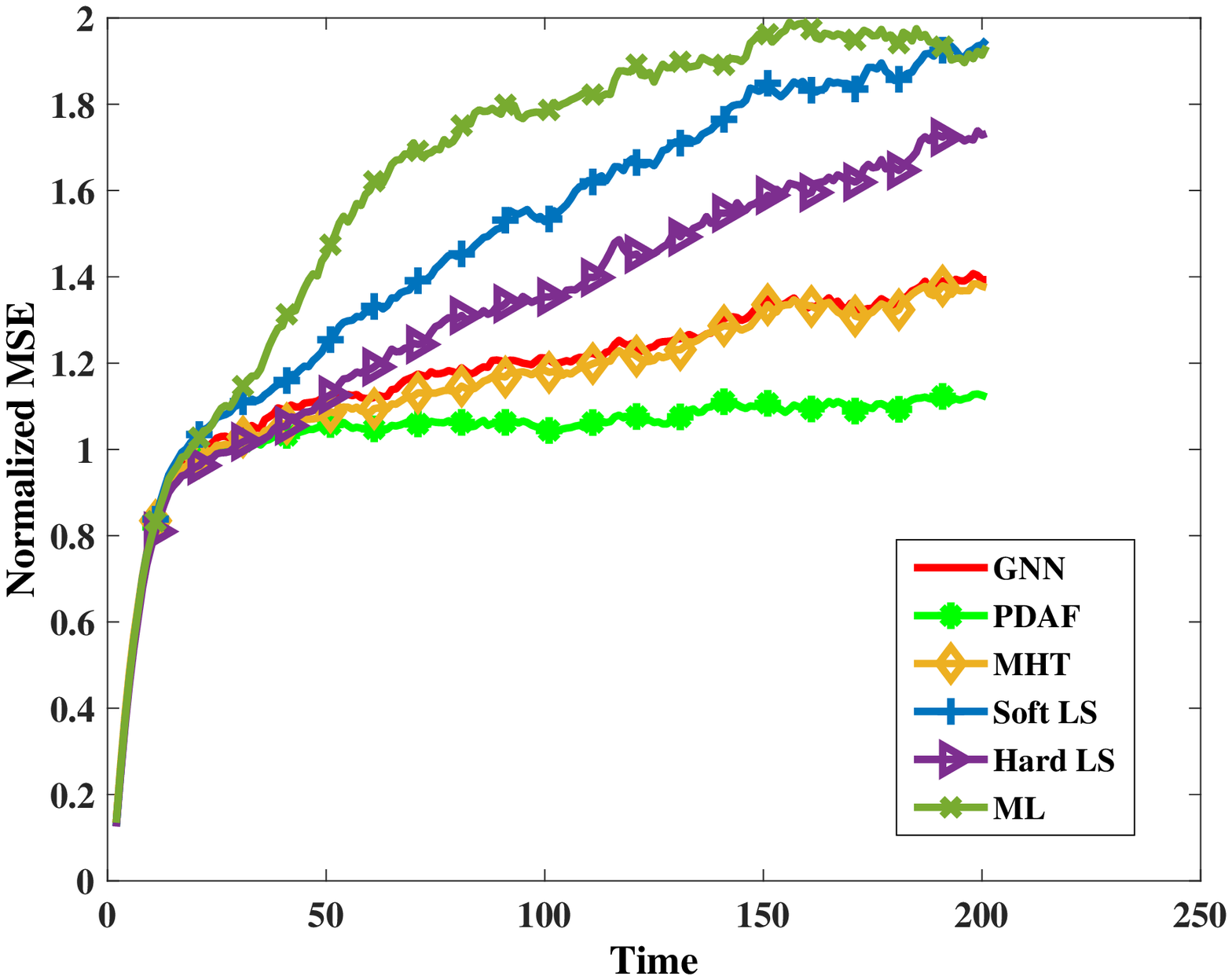}
		\caption{Normalized MSE of uncoordinated trackers with $K=6$, $M=256$.}
		\label{fig:exploitK6M256}
	\end{minipage}
\end{figure}

For $K=6$ devices and $M=16$, $M=256$ antennas, Normalized MSE (NMSE) for all uncoordinated trackers are plotted in \Fref{fig:exploitK6M16} and \Fref{fig:exploitK6M256} respectively. Note that the optimal uncoordinated tracker is too complex to be practical and thus overlooked. As for PoA, it is definitely lower than the best performing uncoordinated algorithm which is suggested to be PDAF by the figures. A major difference emerges between $M=16$ and $M=256$. For $M=16$, ordinary MIMO, NMSE is about $1.5$ for PDAF which means that worst case PoA equals half the coordinated MMSE. The PDAF NMSE for $M=256$ is about $1.1$ which means that worst case PoA is about one tenth of the coordinated MMSE. This result is remarkable in the sense that if we go into the massive MIMO regime, a practical uncoordinated algorithm like PDAF gets very close to the performance of optimum coordinated tracker. However, the gap is considerably larger for ordinary MIMO. Same conclusion is valid for all other uncoordinated trackers as their NMSE is considerably smaller in the massive MIMO regime. \Fref{fig:exploitK6M16} suggests that MHT (4 best hypotheses), GNN, and ML perform slightly better than Soft/Hard LS. For the massive MIMO setup in \Fref{fig:exploitK6M256}, same pattern is observed with the exception of ML which performs the poorest. Note the $y$-axis scale difference in \Fref{fig:exploitK6M16} versus \Fref{fig:exploitK6M256}.

\begin{figure}[t]
	\centering
	
	\begin{minipage}{.5\textwidth}
		\centering
		\captionsetup{font=small, width=.8\linewidth}
		\includegraphics[width=.9\linewidth]{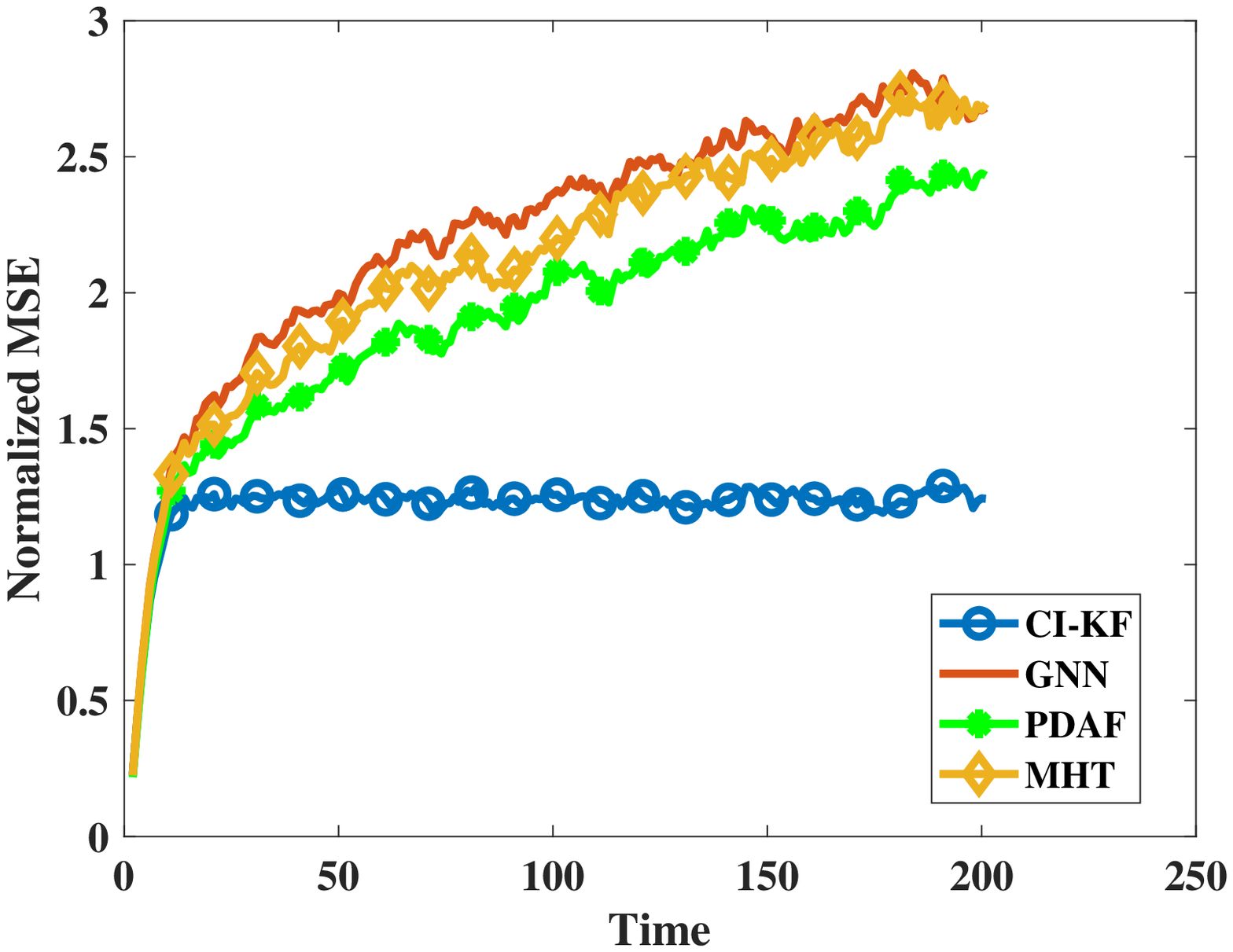}
		\caption{Normalized MSE of uncoordinated trackers when discarding collisions with $K=2$, $M=16$.}
		\label{fig:discardK2M16}
	\end{minipage}%
	\begin{minipage}{.5\textwidth}
		\centering
		\captionsetup{font=small, width=.8\linewidth}
		\includegraphics[width=.9\linewidth]{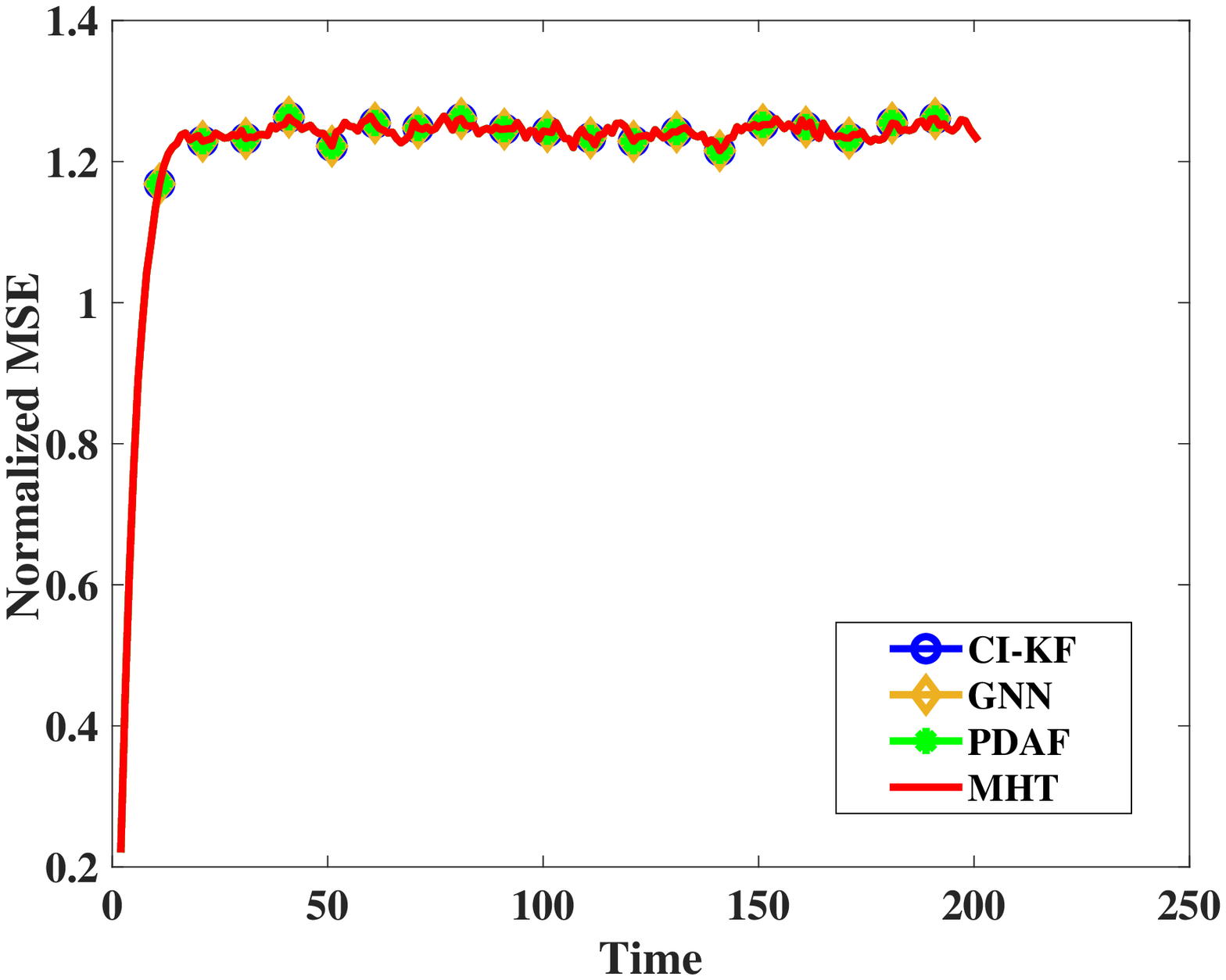}
		\caption{Normalized MSE of uncoordinated trackers when discarding collisions with $K=2$, $M=256$.}
		\label{fig:discardK2M256}
	\end{minipage}
\end{figure}

For the uncoordinated algorithms, if we discard collisions when $K$ is not small, we practically discard all measurements and will keep only predicting the channels for all the users. However, when $K=2$, we obtain reasonable performance with collisions discarded. \Fref{fig:discardK2M16} illustrates that CI-KF NMSE levels off at $1.2$ which means that performance is reduced by 20 percent while we get a complexity improvement by the factor of 4. As for uncoordinated trackers which discard collisions, they are limited by the fact that the estimate \eqref{na} is not accurate in low antenna regime and overestimates the number of active devices predicting many nonexistent collisions. Thus, PDAF, MHT, and GNN only predict most of the time. For massive MIMO, this limitation is no longer a burden as the estimate in \eqref{na} becomes very accurate. Subsequently, PDAF, MHT, and GNN perform similar to the coordinated CI-KF suffering an NMSE of about $1.2$ as witnessed by \Fref{fig:discardK2M256}. Therefore, if we compare optimum coordinated versus uncoordinated trackers which discard collisions, PoA equals zero, which is again a remarkable merit for massive MIMO.

\subsection{Achievable Data Rates}
\begin{figure}[t]
	\centering
	
	\begin{minipage}{.5\textwidth}
		\centering
		\captionsetup{font=small, width=.8\linewidth}
		\includegraphics[width=.9\linewidth]{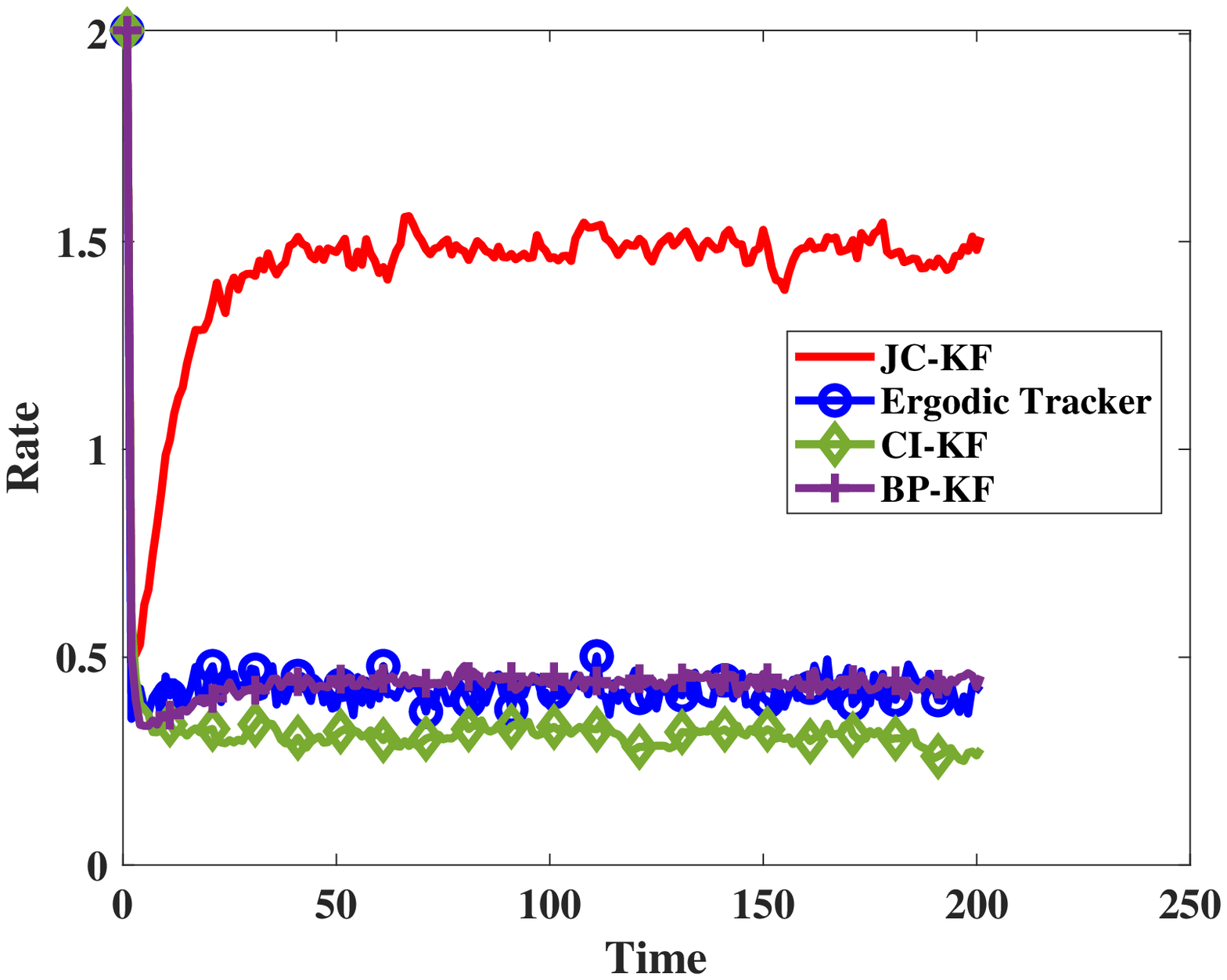}
		\caption{MRT data rate of coordinated trackers with $K=6$, $M=16$.}
		\label{fig:McoordinateK6M16}
	\end{minipage}%
	\begin{minipage}{.5\textwidth}
		\centering
		\captionsetup{font=small, width=.8\linewidth}
		\includegraphics[width=.9\linewidth]{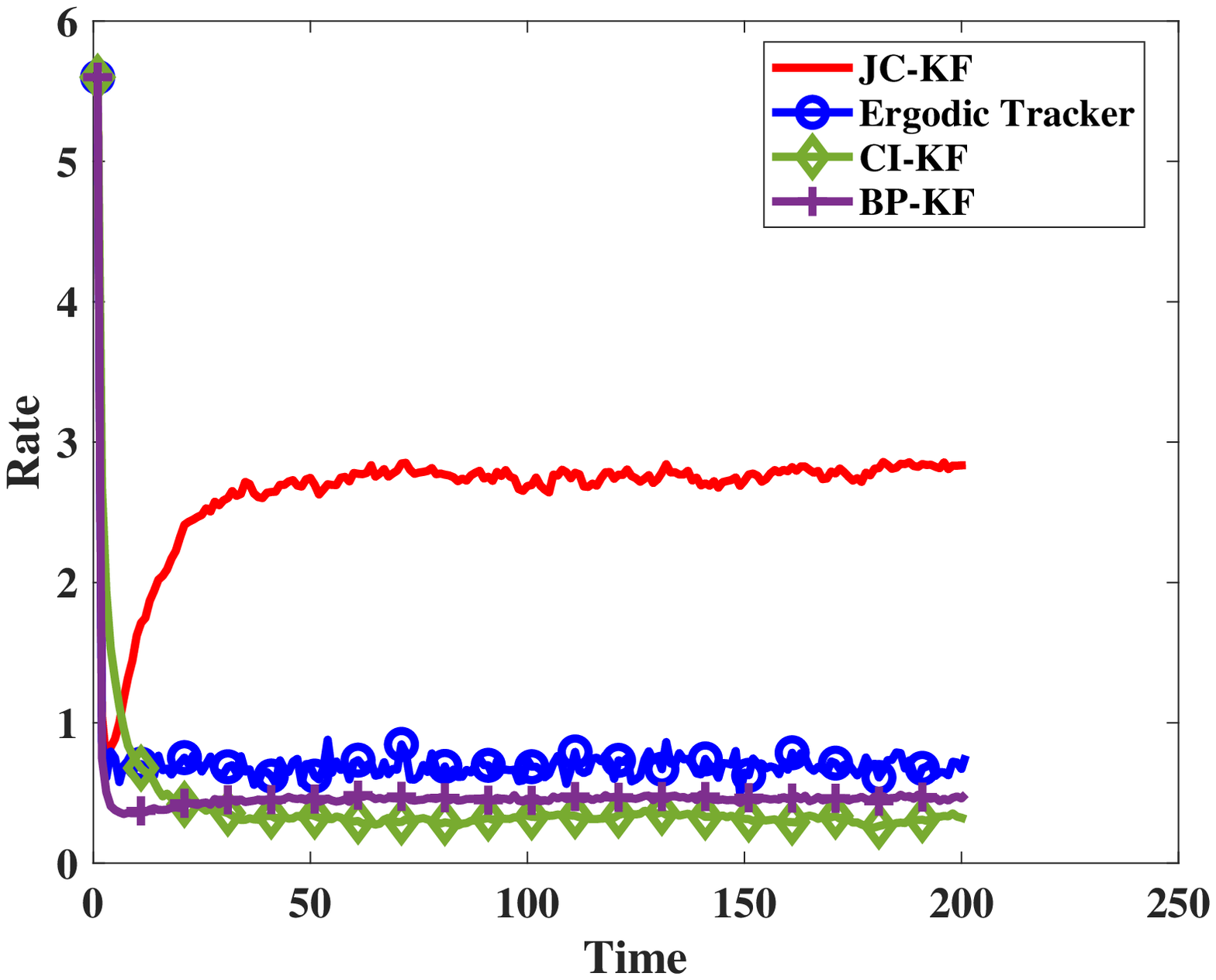}
		\caption{MRT data rate of coordinated trackers with $K=6$, $M=256$.}
		\label{fig:McoordinateK6M256}
	\end{minipage}
\end{figure}

The final aim in any communication system is to endow each individual user with the highest possible data-rate that can be secured with the available CSI. We characterize capacity for user $k$ at time $t$ as 
\begin{equation} \label{capacity}
C_t^k=\log_2\left(1+\text{SINR}\right)=\log_2 \left(1+\frac{|{\bh^{(k)}_{t}}^{T}\bz_{t}^{(k)}|^{2}}{\sum_{\ell\neq k}|{\bh^{(k)}_{t}}^{T}\bz_{t}^{(\ell)}|^{2}+\sigma_{v}^{2}}\right)
\end{equation}
where $\bz_t^{(k)}$ is the linear beamforming weight for user $k$ at time $t$ and $\sigma_v^2$ is the receiver noise variance. First, we use a simple MRT beamformer. \Fref{fig:McoordinateK6M16} and \Fref{fig:McoordinateK6M256} show a comparison of the JC-KF, CI-KF, and BP-KF with respect to the capacity for user one for the cases of $ K=6\hspace{0.1cm},~M=16$ and $ K=6,~M=256$, respectively. It can be observed that joint tracking of all $K$ users does indeed greatly improve the data rate compared to decoupled sub-optimal individual trackers. With $K=6$, which means too many collisions, CI-KF basically predicts all devices channels most of the time and measurements are not exploited at all. Indeed, CI-KF performs dead reckoning whose error increases in the long run. BP-KF performs better than CI-KF but the extra gain is hardly noteworthy. We conclude that more advanced sub-optimal trackers might be needed here to fill the performance gap between JC-KF and BP-KF. In spite of its poor MSE, Ergodic tracker performs similar to CI-KF and BP-KF in terms of data-rate for ordinary MIMO and outperforms them in massive MIMO.  

\begin{figure}[t]
	\centering
	
	\begin{minipage}{.5\textwidth}
		\centering
		\captionsetup{font=small, width=.8\linewidth}
		\includegraphics[width=.9\linewidth]{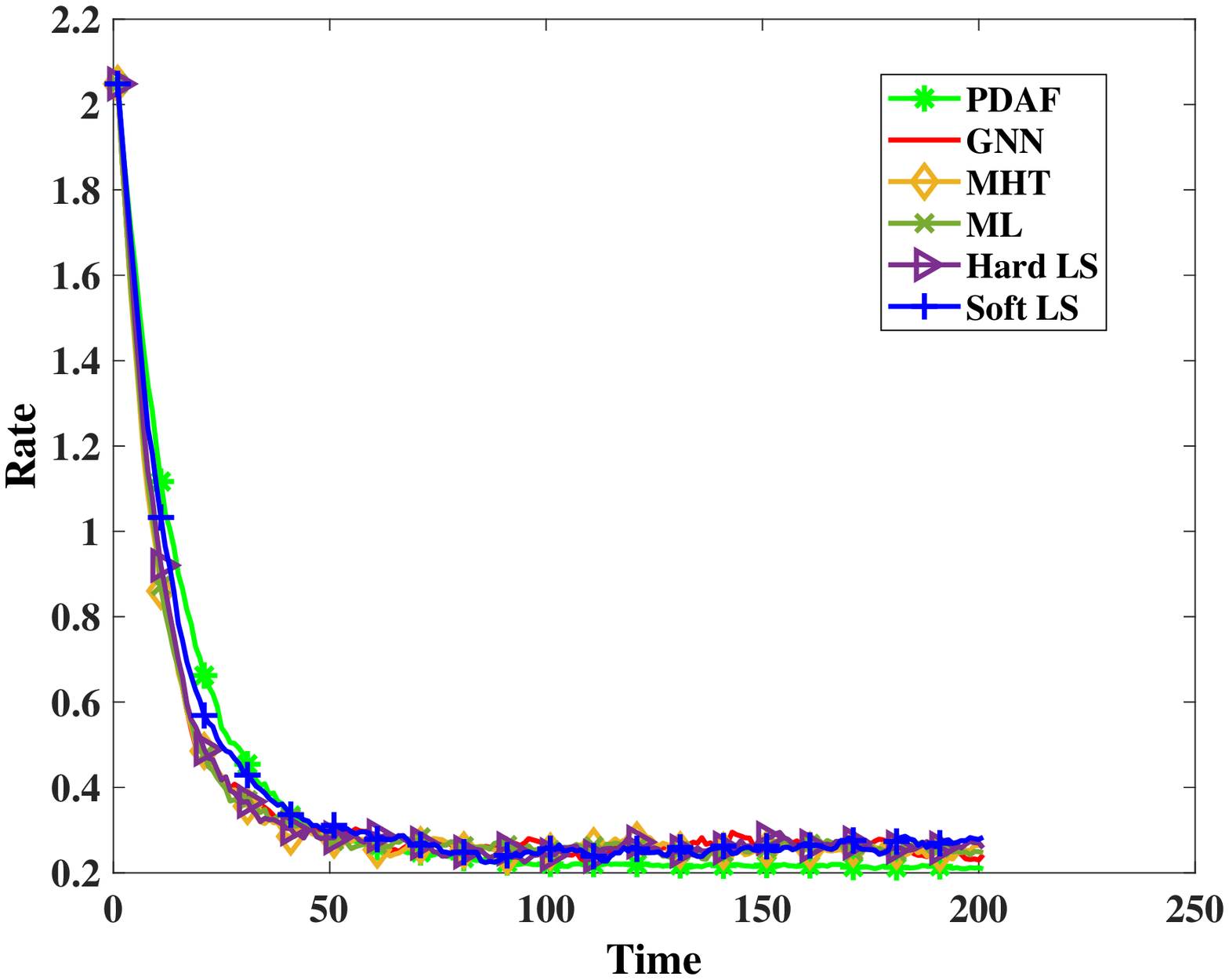}
		\caption{MRT data rate of uncoordinated trackers with $K=6$, $M=16$.}
		\label{fig:MdiscardK6M16}
	\end{minipage}%
	\begin{minipage}{.5\textwidth}
		\centering
		\captionsetup{font=small, width=.8\linewidth}
		\includegraphics[width=.9\linewidth]{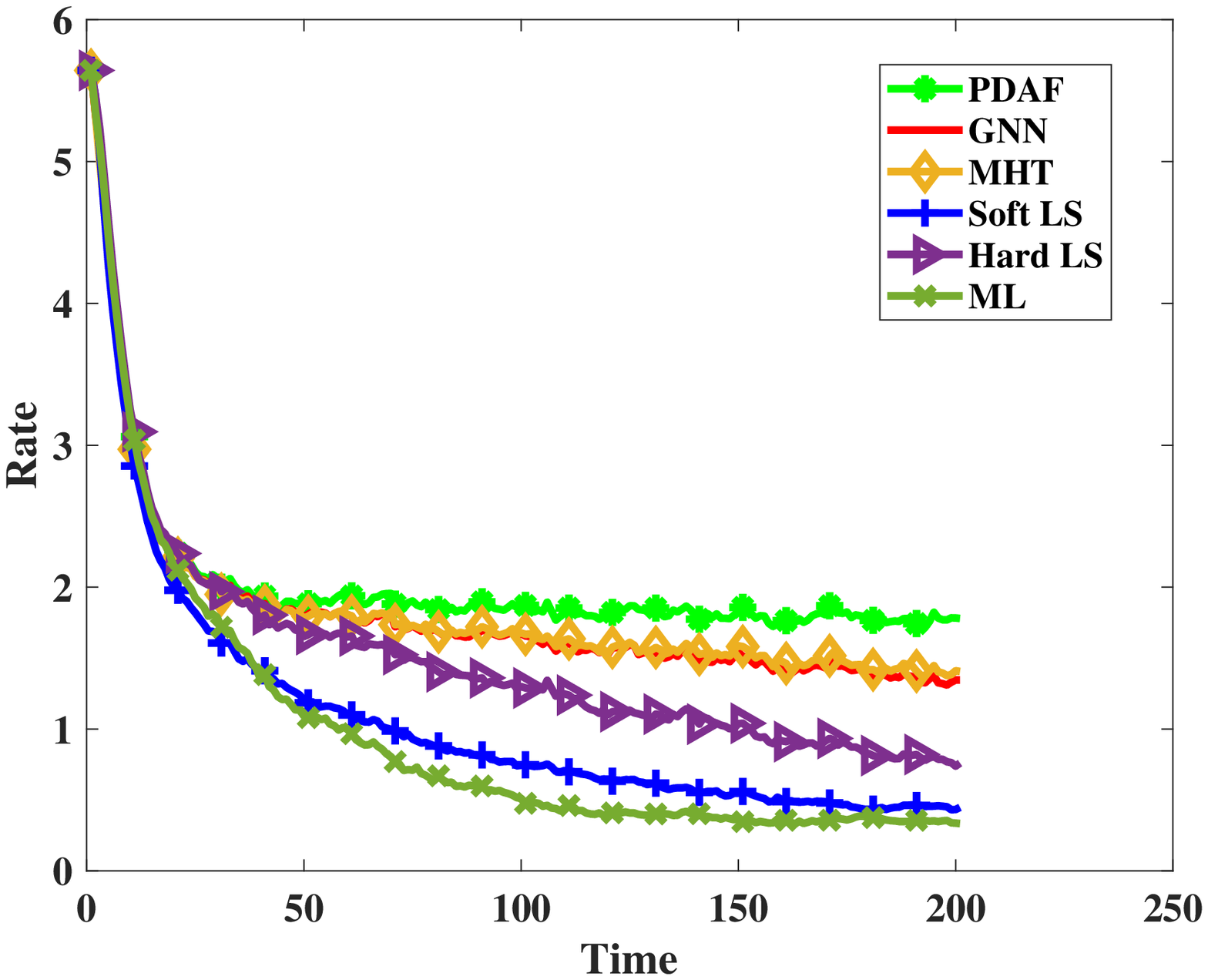}
		\caption{MRT data rate of uncoordinated trackers with $K=6$, $M=256$.}
		\label{fig:MdiscardK6M256}
	\end{minipage}
\end{figure}

MRT beamformer data rates for uncoordinated trackers are plotted in \Fref{fig:MdiscardK6M16} and \Fref{fig:MdiscardK6M256} for $K=6,~M=16$ and $K=6,~M=256$ respectively. It is observed that for ordinary MIMO there is no significant gap between data rates of various algorithms and they all perform poorly. However, in the massive MIMO setting, performance gap is considerable with PDAF performing best and locally optimal ML performing worst. Finally, a robust beamformer that exploits both channel estimates and its covariance matrix was considered \cite{WP09}. This beamformer proved too complex to design for large $K,M$ . When $K=2,~M=16$, capacity of this robust beamformer for various algorithms that discard collisions are plotted in \Fref{fig:RdiscardK2M16}. 

\section{Conclusion}
High-rate IoT setup in massive MIMO was considered and a novel method to track all devices channels simultaneously was advocated. Utilizing a dynamical model for IoT devices channel evolution over time, optimal and various sub-optimal trackers were proposed for coordinated and uncoordinated scenarios. Fundamental performance gap between coordinated and uncoordinated trackers was evaluated analytically. Finally, the performance of various trackers were investigated through extensive simulations.
 
\section*{Appendices}

\section*{Appendix A. Derivation of BP-KF}
\begin{figure}[t]
	\centering
	
	\begin{minipage}{.5\textwidth}
		\centering
		\captionsetup{font=small, width=.8\linewidth}
		\includegraphics[width=.9\linewidth]{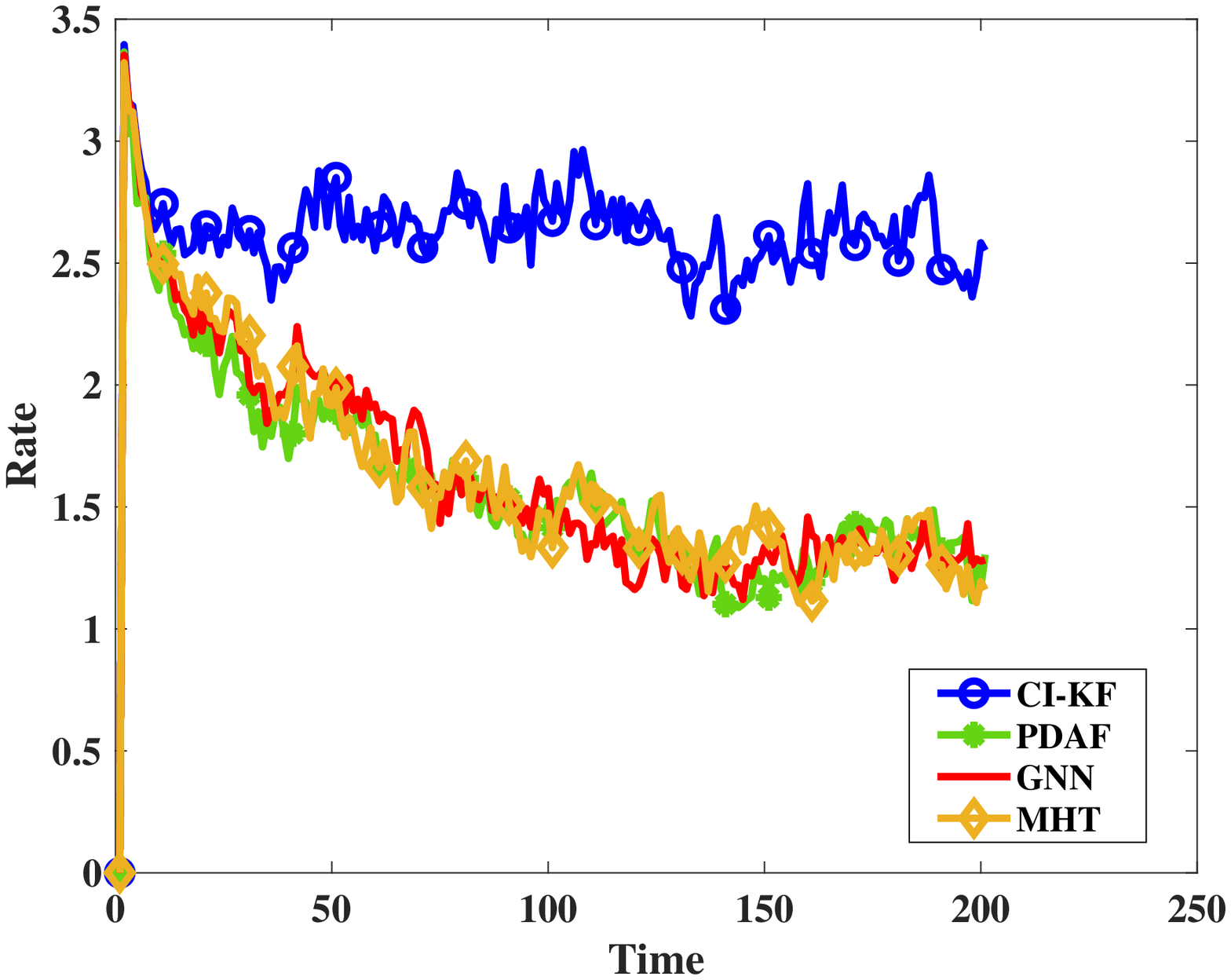}
		\caption{Robust beamformer for various trackers with $K=2$, $M=16$}
		\label{fig:RdiscardK2M16}
	\end{minipage}%
	\begin{minipage}{.5\textwidth}
		\centering
		\captionsetup{font=small, width=.8\linewidth}
		\includegraphics[width=.9\linewidth]{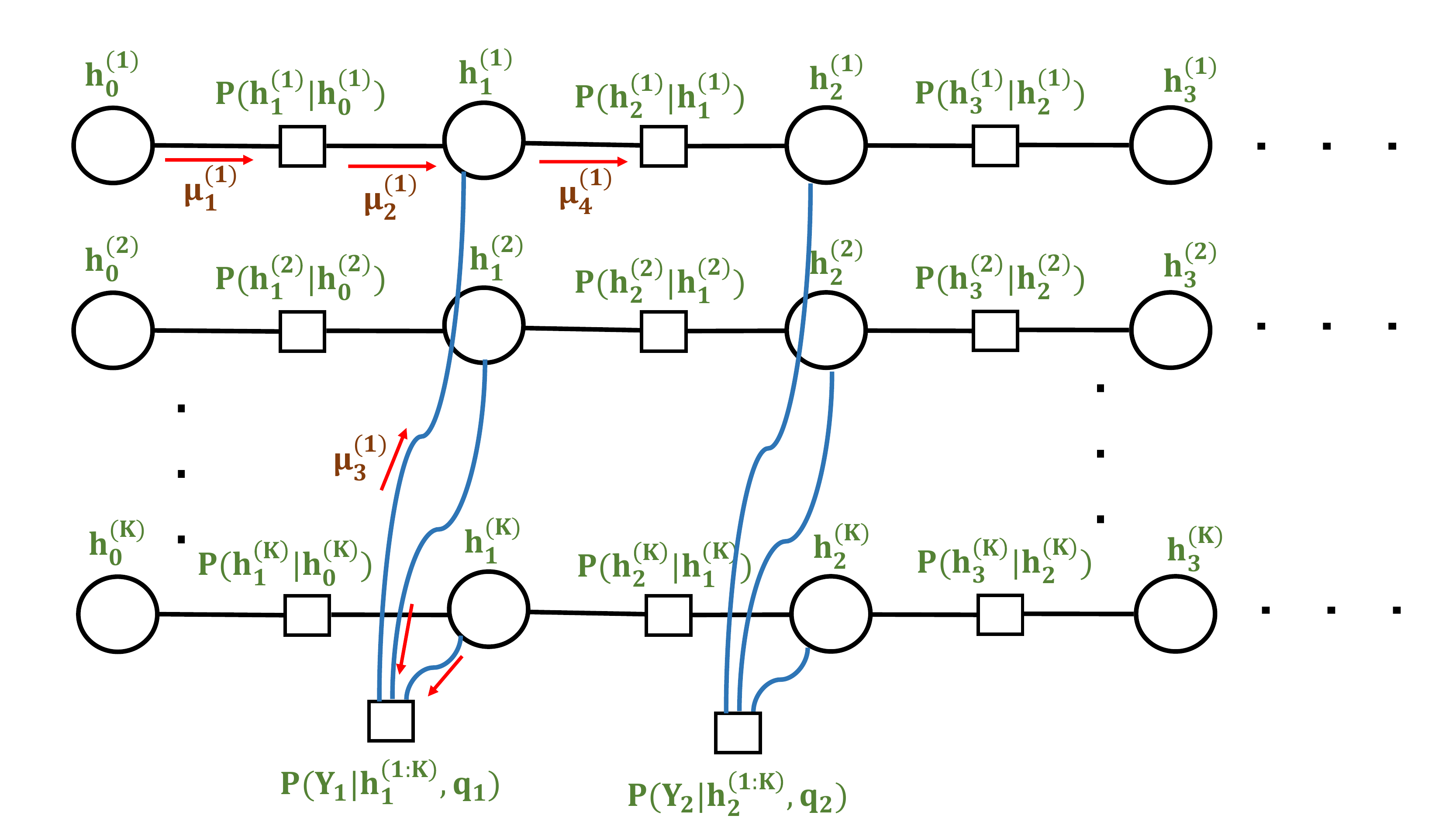}
		\caption{Factor graph for our filtering scenario and belief schedules.}
		\label{fig:BF2}
	\end{minipage}
\end{figure}

To derive BP-KF, we use the factor graph notion as provided in \cite{KFL01}. In their seminal paper, it was shown that KF can alternatively be viewed as belief propagation, or more generally a message passing algorithm on a factor graph. Following the same procedure, in the coordinated setup, factor graph for our scheme is depicted in Fig. \ref{fig:BF2}. 
To proceed, we derive the four messages, or beliefs, denoted as $\mu_1^{(1)},\mu_2^{(1)},\mu_3^{(1)},\mu_4^{(1)}$ in Fig. \ref{fig:BF2}. Then, we generalize these messages to other users and future time slots. Note that given the joint Gaussianity of the overall model, belief messages are Gaussian themselves. Therefore, we only need to include the mean and covariance in the beliefs.
\begin{align*}
\mu_{1}^{(1)}(\bh_0^{(1)}) &= p(\bh_0^{(1)})=\cN(\bh_0^{(1)};\hat{\bh}_{0}^{(1)},\bP_{00}^{(1)})\\
 \mu_2^{(1)}(\bh_1^{(1)}) &= \int p(\bh_1^{(1)}|\bh_0^{(1)})\mu_{1}^{(1)}(\bh_0^{(1)}) ~d\bh_0^{(1)}= \int p(\bh_1^{(1)}|\bh_0^{(1)})p(\bh_0^{(1)}) ~d\bh_0^{(1)}\\ \nonumber
&=\int \cN(\bh_1^{(1)};\bA_1^{(1)}\bh_0^{(1)},\bQ_1^{(1)}) ~\cN(\bh_0^{(1)};\hat{\bh}_{0}^{(1)},\bP_{00}^{(1)})~ d\bh_0^{(1)}\\&=\cN\left(\bh_1^{(1)};\bA_1^{(1)}\hat{\bh}_0^{(1)},\bA_1^{(1)}\bP_{00}^{(1)}\bA_1^{(1)^T}+\bQ_1^{(1)}\right):=\cN\left(\bh_1^{(1)};\hat{\bh}_{1|0}^{(1)},\bP_{10}^{(1)}\right) 
\end{align*}
These beliefs are exactly same as those for ordinary KF. In evaluating $\mu_3^{(1)},\mu_4^{(1)}$ differences appear:
\begin{align*}
 \mu_3^{(1)}(\bh_1^{(1)})&=\int p(\by_1|\bh_1,\bq_1) \prod_{j: q_1^{(j)}=1,~j\neq 1,} ~\mu_2^{(j)}(\bh_1^{(j)}) ~~~d \bh_1^{(j)} \\
& = \int \cN\left(\by_1;\sum_{k=1}^K q_1^{(k)}\bh_{1}^{(k)},\bR_1\right) \prod_{j: q_1^{(j)}=1,~j\neq 1,} ~ \cN\left(\bh_1^{(j)};\hat{\bh}_{1|0}^{(j)},\bP_{10}^{(j)}\right)~~~d \bh_1^{(j)} \\
& =\cN\left(\by_1-\sum_{j=2}^K q_1^{(j)}\hat{\bh}_{1|0}^{(j)};\bh_1^{(1)},\bR_1+\sum_{j=2}^K q_1^{(j)}\hat{\bP}_{1|0}^{(j)}\right):=\cN\left(\hat{\by}_1^{(1)};\bh_1^{(1)},\hat{\bR}_1^{(1)}\right)
\end{align*}
The above expression is valid only if $q_1^{(1)}=1$, that is user $1$ participates in the collision. Otherwise, there is no connection between variable node $\bh_1^{(1)}$ and factor node $p(\by_1|\bh_1,\bq_1)$ and there will be no $\mu_3^{(1)}$ and we will have $\mu_4^{(1)}=\mu_2^{(1)}$. Finally, $\mu_4^{(1)}$ is computed as in an ordinary KF.
\begin{align*}
\mu_4^{(1)}(\bh_1^{(1)})&= \mu_2^{(1)}(\bh_1^{(1)})\mu_3^{(1)}(\bh_1^{(1)})=\cN\left(\bh_1^{(1)};\hat{\bh}_{1|0}^{(1)},\bP_{10}^{(1)}\right)\cN\left(\hat{\by}_1^{(1)};\bh_1^{(1)},\hat{\bR}_1^{(1)}\right)=\cN\left(\bh_1^{(1)};\hat{\bh}_{1|1}^{(1)},\bP_{1|1}^{(1)}\right)
\end{align*}
where the parameters on the last line are given by \eqref{mybp}. Same derivation can be extended in a straightforward manner to other users and future time slots which completes the proof.
\section*{Appendix B. Proof of Theorem 1}

\begin{align}
\label{eq:43}
&\hspace{-4cm}\text{PoA}=\mathbb{E}_{\by_{1}}\left[\sum_{\bq_{1}}~p(\bq_{1}|\by_{1})\|\hat{\bh}_{1|1}(\bq_{1})-\hat{\bh}^{(u)}_{1|1}\|^2\right]\nonumber\\&\hspace{-3.3cm}=
\mathbb{E}\left[\sum_{{\bf{q}}_{1}}p({\bf{q}}_{1}|{\bf{y}}_{1}) \left( \| \hat{{\bf{h}}}_{1|1}({\bf{q}}_{1})\|^{2}+\| \hat{{\bf{h}}}_{1|1}\|^{2}-2 \hat{{\bf{h}}}_{1|1}({\bf{q}}_{1})\hat{{\bf{h}}}_{1|1}\right)\right]\nonumber \\&\hspace{-3.3cm}=\mathbb{E}\left[\sum_{\bq_{1}}p({\bf{q}}_{1}|{\bf{y}}_{1}) \left(\| \hat{{\bf{h}}}_{1|1}({\bf{q}}_{1})\|^{2}+\| \hat{{\bf{h}}}_{1|1}\|^{2} \right) \right]
\end{align}
Note that $\hat{{\bf{h}}}_{1|1}({\bf{q}}_{1})$ amounts to a correction step assuming $\bq_1$ is the true model, while $\hat{{\bf{h}}}_{1|1}$ is the soft combination of all possible $\bq_1$ as derived for the optimal uncoordinated tracker. They can further be written as
\begin{align}
&\hat{{\bf{h}}}_{1|1}({\bf{q}}_{t})=\hat{{\bf{h}}}_{1|0}+{\bf{K}}({{\bf{q}}_{1}})\left({\bf{y}}_{1}-{\bf{B}}({{\bf{q}}}_{1})\hat{{\bf{h}}}_{1|0}\right)\\ \nonumber
&\hat{{\bf{h}}}_{1|1}=\sum_{\bq_{1}}p({\bf{q}}_{1}|{\bf{y}}_{1}) \left[\hat{{\bf{h}}}_{1|0}+{\bf{K}}({\bf{q}_{1}})\left({\bf{y}}_{1:t}-{\bf{B}}({\bf{q}}_{1})\hat{{\bf{h}}}_{1|0}\right) \right]\\ \nonumber
&\hspace{0.9cm}=\hat{{\bf{h}}}_{1|0}+\sum_{\bq_{1}}p({\bf{q}}_{1}|{\bf{y}}_{1}){{\bf{K}}}({\bf{q}}_{1})\left({\bf{y}}_{1}-{{\bf{B}}}({{\bf{q}}_{1}})\hat{\bf{h}}_{1|0}\right)
\end{align}
Plugging into PoA in \eqref{eq:43}, we get
\begin{align}
 \nonumber
&\hspace{-0.7cm}\text{PoA}=\mathbb{E}\left[ \sum_{\bq_{1}}p({\bf{q}}_{t}|{\bf{y}}_{1}) \left(\hat{{\bf{h}}}_{1|0}^{T}{{\bf{K}}}({{\bf{q}}_{1}})\big({\bf{y}}_{1}-{{\bf{B}}}({{\bf{q}}_{1}})\hat{{\bf{h}}}_{1|0}\big)\right) \right]\\ \nonumber
&\hspace{0.3cm}+\mathbb{E}\left[  \sum_{\bq_{t}}p({\bf{q}}_{1}|{\bf{y}}_{1}) \left \| {{\bf{K}}}({{\bf{q}}_{1}})\left({\bf{y}}_{1}-{{\bf{B}}}({{\bf{q}}_{1}})\hat{{\bf{h}}}_{1|0}\right) \right\|^{2}\right] \\ \nonumber
& -\mathbb{E}\left[\sum_{\bq_{1}}p({\bf{q}}_{1}|{\bf{y}}_{1})\hat{{\bf{h}}}_{1|0}^{T}{{\bf{K}}}({{\bf{q}}_{1}})\left({\bf{y}}_{1}-{{\bf{B}}}({{\bf{q}}_{1}})\hat{{\bf{h}}}_{1|0}\right)\right]
\\ \nonumber
&\hspace{0.3cm}-\mathbb{E}\left[\left \|\sum_{\bq_{1}}p({\bf{q}}_{1}|{\bf{y}}_{1}) {{\bf{K}}}({{\bf{q}}_{1}}) \left({\bf{y}}_{1}-{{\bf{B}}}({{\bf{q}}_{1}})\hat{{\bf{h}}}_{1|0}\right) \right \|^{2} \right] \\ \nonumber
 &\hspace{-0.1cm}=\mathbb{E}\left[  \sum_{\bq_{1}}p({\bf{q}}_{1}|{\bf{y}}_{1})\left \|{{\bf{K}}}({{\bf{q}}_{1}})\left({\bf{y}}_{1}-{{\bf{B}}}({{\bf{q}}_{1}})\hat{{\bf{h}}}_{1|0}\right) \right\|^{2}\right]
 \\ \label{eq:130}
& \hspace{0.1cm}- \mathbb{E}\left[\left \| \sum_{\bq_{1}}p({\bf{q}}_{1}|{\bf{y}}_{1}){{\bf{K}}} ({{\bf{q}}_{1}})\left({\bf{y}}_{1}-{{\bf{B}}}({{\bf{q}}_{1}})\hat{{\bf{h}}}_{1|0}\right)\right \|^{2} \right]
\end{align}

Note that $\|.\|^{2}$ is convex, $(\frac{\partial \|X\|^{2}}{\partial X^{2}}=2I)$. Hence, jensen inequality ensure that \eq{eq:130} is positive since it is equal to the expectation of positive entity ($\|\sum_{i} \lambda_{i} X_{i}\|^{2} \leq \sum_{i} \lambda_{i}\| X_{i}\|^{2}$).

First term in \eq{eq:130} yields
\begin{align}
&\mathbb{E}\left[  \sum_{\bq_{1}}p({\bf{q}}_{1}|{\bf{y}}_{1})\left \|{{\bf{K}}}({{\bf{q}}_{1}})\left({\bf{y}}_{1}-{{\bf{B}}}({{\bf{q}}_{1}})\hat{{\bf{h}}}_{1|0}\right) \right\|^{2}\right]=
  \nonumber \\
&\mathbb{E}\bigg [ \sum_{\bq_{1}}p({\bf{q}}_{1}|{\bf{y}}_{1}) \bigg({\bf{y}}_{1}^{T}{{\bf{K}}}^{T}({{\bf{q}}_{1}}){{\bf{K}}}({{\bf{q}}_{1}}){\bf{y}}_{1}
+\hat{{\bf{h}}}_{1|0}^{T}{{\bf{B}}}({{\bf{q}}_{1}}^{T}){{\bf{K}}}({{\bf{q}}_{1}}^{T}){{\bf{K}}}({{\bf{q}}_{1}})\hat{{\bf{h}}}_{1|0}{{\bf{B}}}({{\bf{q}}_{1}})
  \nonumber \\
&\hspace{3.5cm}-2\hat{{\bf{h}}}_{1|0}{{\bf{B}}}({{\bf{q}}_{1}}^{T}){{\bf{K}}}({{\bf{q}}_{1}})^{T}{{\bf{K}}}({{\bf{q}}_{1}})\bigg) \bigg]
\end{align}
Now, let us evaluate each term independently.
\begin{align}
\label{eq:46}
& \mathbb{E}\left[ p({\bf{q}}_{1}|{\bf{y}}_{1})\right]=\int p(\bq_1|\by_1)~p(\by_1)d\by_1=P({\bf{q}}_{1}) \\  \nonumber
&\mathbb{E}\left[ p({\bf{q}}_{1}|{\bf{y}}_{1}){\bf{y}}_{1} \right]=\int {\bf{y}}_{1}~p({\bf{y}}_{1}|{\bf{q}}_{1})~p({\bf{q}}_{1}) ~~d{\bf{y}}_{1}\\  \nonumber
&\hspace{2.9cm}=\int \int {\bf{y}}_{1}~p({\bf{y}}_{1}|{\bf{h}}_{1}, {\bf{q}}_{1})~p({\bf{h}}_{1}|{\bf{q}}_{1})~p({\bf{q}}_{1})~~d{\bf{y}}_{1}~~d{\bf{h}}_{1}\\  \nonumber
&\hspace{2.9cm}=p({\bf{q}}_{1})~ \int p({\bf{h}}_{1}) ~\int {\bf{y}}_{1}~\cN\left({\bf{y}}_{1};{{\bf{B}}}({{\bf{q}}_{1}}){\bf{h}}_{1},{\bf{R}_1}\right)~~d{\bf{y}}_{1}~~d{\bf{h}}_{1}\\  \nonumber
&\hspace{2.9cm}=p({\bf{q}}_{1}) ~\int {{\bf{B}}}({{\bf{q}}_{1}}){\bf{h}}_{1}p({\bf{h}}_{1})~~d{\bf{h}}_{1}=p({\bf{q}}_{1})~{{\bf{B}}}({{\bf{q}}_{1}})~\mathbb{E}\left[{\bf{h}}_{1}\right]=0 
\end{align}
The last line is zero because $\mathbb{E}\left[{\bf{h}}_{1}\right]=\mathbb{E}\left[{\bf{A}}{\bf{h}}_{0}+{\bf{u}}_{1}\right]=0$. Let us focus on the second moment now.
\begin{align}
&\mathbb{E}\left[p({\bf{q}}_{1}|{\bf{y}}_{1}){\bf{y}}_{1}{\bf{y}}^{T}_{1}\right]=p({\bf{q}}_{1}) \int p({\bf{h}}_{1}) \int {\bf{y}}_{1}~{\bf{y}}_{1}^{T}~\cN\left({\bf{y}}_{1};{{\bf{B}}}({{\bf{q}}_{1}}){\bf{h}}_{1},{\bf{R}_1}\right)~d{\bf{y}}_{1}~d{\bf{h}}_{1}\\  \nonumber
&\hspace{3.6cm}=p({\bf{q}}_{1}) \int p({\bf{h}}_{1}) \left[{{\bf{B}}}({{\bf{q}}_{1}})~{\bf{h}}_{1}{\bf{h}}_{1}^{T}~{{\bf{B}}}({{\bf{q}}_{1}})^{T}+{\bf{R}}\right]~~d{\bf{h}}_{1}\\ \nonumber
&\hspace{3.6cm}=p({\bf{q}}_{1})~ \left[{\bf{R}}+{{\bf{B}}}({{\bf{q}}_{1}})~\mathbb{E}\left[{\bf{h}}_{1}{\bf{h}}_{1}^{T}\right]~{{\bf{B}}}({{\bf{q}}_{1}})^{T} \right]\\  \nonumber
&\mathbb{E}\left[{\bf{h}}_{1}{\bf{h}}_{1}^{T}\right]=\mathbb{E}\left[({\bf{A}}{\bf{h}}_{0}+{\bf{u}}_{1})({\bf{A}}{\bf{h}}_{0}+{\bf{u}}_{1})^{T}\right]={\bf{A}}{\bf{P}}_{0}{\bf{A}}^{T}+{\bf{Q}}
\end{align}
Combining these results, the first term in \eq{eq:130} can be written as follows
\begin{align*}
\label{eq:47}
& \hspace{-0.7cm}\mathbb{E}\left[  \sum_{\bq_{1}}p({\bf{q}}_{1}|{\bf{y}}_{1})\left \|{{\bf{K}}}({{\bf{q}}_{1}})\left({\bf{y}}_{1}-{{\bf{B}}}({{\bf{q}}_{1}})\hat{{\bf{h}}}_{1|0}\right) \right\|^{2}\right]
\\ \nonumber
&\hspace{3.5cm}=\sum_{\bq_{1}} p({\bf{q}}_{1})\hat{{\bf{h}}}_{0}{\bf{A}}^{T}{{\bf{B}}}({{\bf{q}}_{1}}^{T}){{\bf{K}}}({{\bf{q}}_{1}}^{T})\times {{\bf{K}}}({{\bf{q}}_{1}}){{\bf{B}}}({{\bf{q}}_{1}}){\bf{A}}\hat{{\bf{h}}}_{0}\\ \nonumber
&\hspace{3.8cm}+\sum_{\bq_{1}} p({\bf{q}}_{1}) \mbox{trace}\left[{\bf{R}}+{{\bf{B}}}({{\bf{q}}_{t}})\left({\bf{A}}{\bf{P}}_{t-1}{\bf{A}}^{T}+{\bf{Q}}\right){{\bf{B}}}({{\bf{q}}_{1}})^{T}\right]
\end{align*}
Second term in \eq{eq:130} can be simplified as follows
\begin{align*}
&\mathbb{E}\left[\left \| \sum_{\bq_{t}}p({\bf{q}}_{1}|{\bf{y}}_{1}){{\bf{K}}}({{\bf{q}}_{1}})\left({\bf{y}}_{1}-{{\bf{B}}}({{\bf{q}}_{1}})\hat{{\bf{h}}}_{1|0}\right)\right \|^{2}\right] \geq \left\| \mathbb{E}\left[\sum_{\bq_{1}}p({\bf{q}}_{1}|{\bf{y}}_{1}){{\bf{K}}}({{\bf{q}}_{1}})\left({\bf{y}}_{1}-{{\bf{B}}}({{\bf{q}}_{1}})\hat{{\bf{h}}}_{1|0}\right)\right] \right\|^{2}
\\ \nonumber
&\hspace{8.5cm} =\left \| \sum_{\bq_{1}}p({\bf{q}}_{1}) {{\bf{K}}}({{\bf{q}}_{1}}) {{\bf{B}}}({{\bf{q}}_{1}})\hat{{\bf{h}}}_{1|0} \right\|^{2}
\\ \nonumber
 & \hspace{6.2cm}=\sum_{\bq_{1}} \sum_{\tilde{\bq}_{1}} p({\bf{q}}_{1})p({\tilde{\bf{q}}}_{1})\hat{{\bf{h}}}_{1|0}^{T}{{\bf{B}}}({{\bf{q}}_{1}})^{T}{{\bf{K}}}({{\bf{q}}_{1}})^{T}{{\bf{K}}}({{\tilde{\bf{q}}}_{1}}){{\bf{B}}}({{\tilde{\bf{q}}}_{1}})\hat{{\bf{h}}}_{1|0}
 \\ \nonumber
 &\hspace{6.2cm}=\sum_{\bq_{1}} \sum_{\tilde{\bq}_{1}} p({\bf{q}}_{1})p({\tilde{\bf{q}}}_{1})\hat{{\bf{h}}}_{0}^{T}{\bf{A}}^{T}{{\bf{B}}}({{\bf{q}}_{1}})^{T}{{\bf{K}}}({{\bf{q}}_{1}})^{T}{{\bf{K}}}({{\tilde{\bf{q}}}_{1}}){{\bf{B}}}({{\tilde{\bf{q}}}_{1}}){\bf{A}}\hat{{\bf{h}}}_{0}
\end{align*}
We used Jensen inequality on the first line. Plugging back into \eqref{eq:130} completes the proof.




\end{document}